\documentclass[11pt,twoside,openright]{book}
\usepackage{amsfonts}
\usepackage{multicol}
\usepackage{iws2013style}
\usepackage{pdfpages}
\makeatletter
\def\cleardoublepage {\clearpage \if@twoside
\ifodd \c@page
\else \hbox {}\thispagestyle{empty}\newpage
\if@twocolumn \hbox {}\newpage \fi\fi\fi}
\makeatother

\usepackage[latin1]{inputenc}
\usepackage[tbtags]{amsmath}
\usepackage{amssymb}
\usepackage{color}
\usepackage{graphicx}
\usepackage[round]{natbib}
\usepackage{amsbsy}
\usepackage{enumerate}
\usepackage{bm}
\usepackage{moreverb}
\usepackage{multirow}
\usepackage{dcolumn}
\usepackage[hang,bf]{caption}
\usepackage{longtable}
\usepackage{eurosym}

\allowdisplaybreaks[4] %allow breaks of multiline formulas

\numberwithin{equation}{section}  %define numbering depth of equations
\newcommand{\Y}{\ensuremath{\mathcal{Y}}}

\newcommand{\ess}{\ensuremath{\text{ESS}}}
\newcommand{\parms}{\bm{\vartheta}}

\newcommand{\Tb}{\ensuremath{\mathbf{T}}}  %T bold
\newcommand{\db}{\ensuremath{\mathbf{d}}}  %d bold
\newcommand{\zb}{\ensuremath{\mathbf{z}}}  %z bold
\newcommand{\xb}{\ensuremath{\mathbf{x}}}  %x bold
\newcommand{\yb}{\ensuremath{\mathbf{y}}}  %y bold
\newcommand{\Rb}{\ensuremath{\mathbf{R}}}  %R bold
\newcommand{\Lb}{\ensuremath{\mathbf{L}}}  %L bold
\newcommand{\Mb}{\ensuremath{\mathbf{M}}}  %M bold
\newcommand{\Kb}{\ensuremath{\mathbf{K}}}  %K bold
\newcommand{\Db}{\ensuremath{\mathbf{D}}}  %D bold
\newcommand{\Ib}{\ensuremath{\mathbf{I}}}  %I bold
\newcommand{\thetab}{\ensuremath{\bm{\theta}}}  %theta bold
\newcommand{\yos}{\ensuremath{\mathbf{y}_{1:s}}} %y from one to s
\newcommand{\xios}{\ensuremath{\bm{\xi}_{1:s}}} %xi from one to s
\newcommand{\xoM}{\ensuremath{\mathbf{x}_{1:M}}}  %x from 1 to M
\begin{document}
\includepdf[pages={1}]{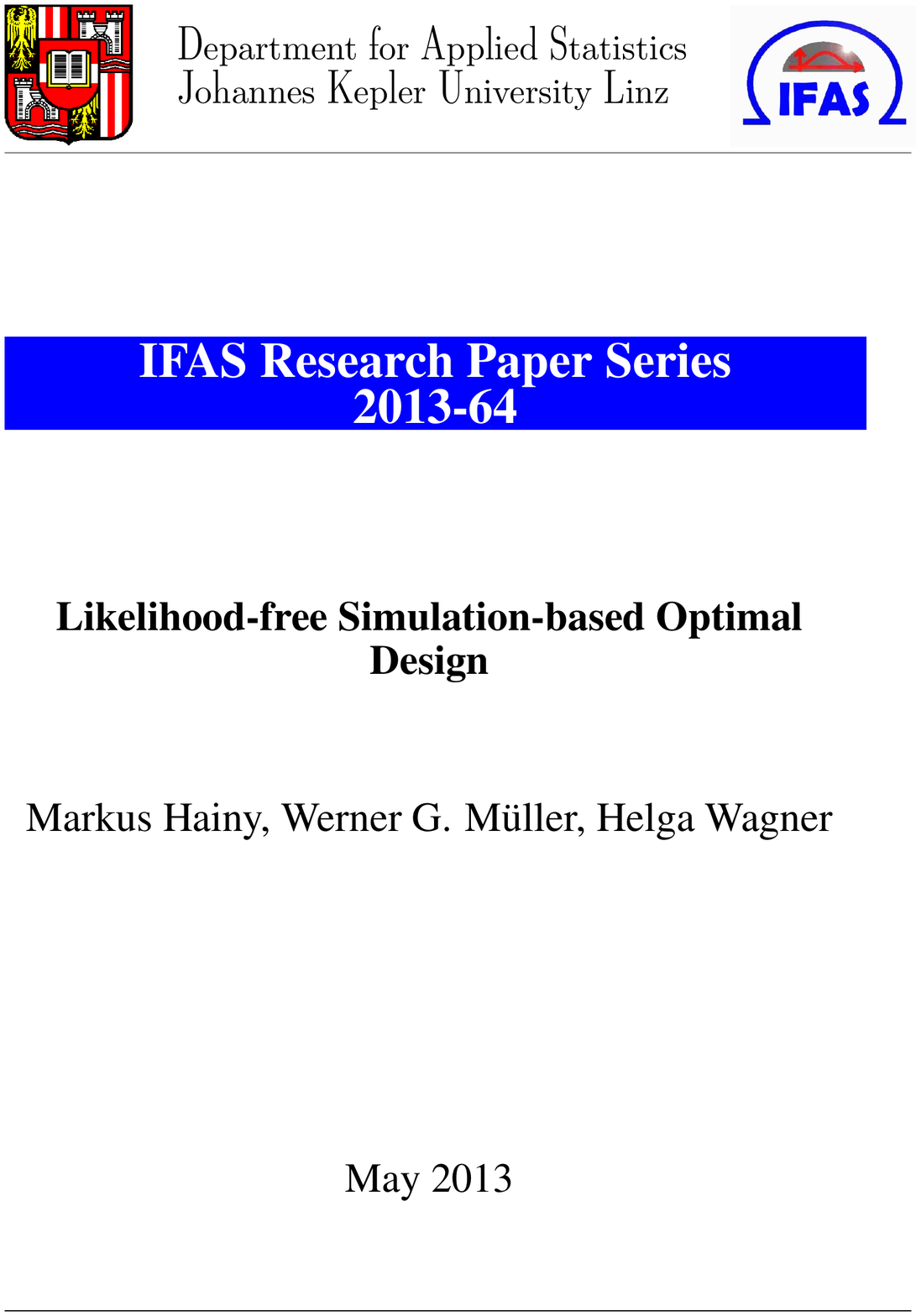}

%{\centering{\parbox[t]{122.5mm}{\protect \centering
%
%\articletitle{ Likelihood-free Simulation-based Optimal Design }}
%
%\vskip 26.4 true pt
%
%{\Size{10}{12}{Markus Hainy, Werner G\@. M\"{u}ller, Helga Wagner\\}}
%{\Size{10}{12}{Johannes Kepler University Linz \\ \texttt{markus.hainy@jku.at, werner.mueller@jku.at, helga.wagner@jku.at}\varspace}}
%
%%% For Authors belonging to the same Institute
%
%{\Size{10}{12}{Werner G\@. M\"{u}ller, Helga Wagner\\}}
%{\Size{10}{12}{Johannes Kepler University Linz\\ \texttt{werner.mueller@jku.at, helga.wagner@jku.at}\varspace}}
%}}
%\vskip 26.4 true pt

\noindent{\bf Abstract.} Simulation-based optimal design techniques are a convenient tool for solving a particular class of optimal design problems. The goal is to find the optimal configuration of factor settings with respect to an expected utility criterion. This criterion depends on the specified probability model for the data and on the assumed prior distribution for the model parameters. We develop new simulation-based optimal design methods which incorporate likelihood-free approaches and utilize them in novel applications.

Most simulation-based design strategies solve the intractable expected utility integral at a specific design point by using Monte Carlo simulations from the probability model. Optimizing the criterion over the design points is carried out in a separate step. \citet{mue:sim} introduces an MCMC algorithm which simultaneously addresses the simulation as well as the optimization problem. In principle, the optimal design can be found by detecting the  utility mode of the sampled design points. Several improvements have been suggested to facilitate this task for multidimensional design problems (see e\@.g\@. \citealt{amz-etal:int}).

We aim to extend this simulation-based design methodology to design problems where the likelihood of the probability model is of an unknown analytical form but it is possible to simulate from the probability model. We further assume that prior observations are available. In such a setting it is seems natural to employ approximate Bayesian computation (ABC) techniques in order to be able to simulate from the conditional probability model. We provide a thorough review of adjacent literature and we investigate the benefits and the limitations of our design methodology for a particular paradigmatic example.
\vskip 26.4  true pt


\section{Introduction}

In the past decades simulation techniques, particularly the use of Markov chain Monte Carlo methods, have revolutionized statistical inference (cf. \citealt{rob-cas:mon}). There has, however, been little impact of this revolution on the experimental design literature other than the pioneering work initiated by Peter M\"uller (cf. \citealt{mue:sim} and \citealt{mue-etal:inh}) and his followers. The main reason may be that M\"uller's method has been limited to a specific choice of design criteria and has effectively been applicable only for finding rather small size designs. Another drawback has been the necessity to be able to explicitly specify the likelihood function for the problem, which can be an obstacle for complex settings faced in real applications.

With the advent of so-called likelihood-free (or approximate Bayesian computation - ABC) methods, the latter issue can be overcome, and we therefore propose to employ those techniques also for finding optimal experimental designs. There are essentially two ways of accomplishing this: the first is to marry ABC with M\"uller's essentially MCMC-based methods, the second is a more integrated albeit more complex implementation avoiding the MCMC step. While the latter has been put forward in \citet{hai-etal:abc}, the former was alluded to in \citet{hai-etal:con} and forms the essence of the present report.

Here we first in Section 2 give a thorough review of the essentials of simulation-based optimal design as well as the various improvements and modifications lately suggested. This ends with a particular emphasis on sequential Monte Carlo methods, which will eventually lend themselves naturally for solving adaptive design problems. Section 3 then presents modifications for including prior observations for the static and sequential design setup. Our main contribution is in Section 4, where we first review ABC methods that are relevant for our purposes and then suggest two MCMC-based optimal design procedures building upon them. Section 5 eventually discusses the type of design criteria / utility functions we can encompass and related implementation issues. Finally in Section 6 we illustrate our methods on a toy example which can be easily related to and understood from classic optimal design theory.

Note that shortly before finalizing this report we have learned of the unpublished paper by \citet{dro-pet:int}, wherein similar ideas have been developed independently. However, while the basic concept of fusing M\"uller's MCMC-based methods with ABC is essentially the same, our approach differs in various ways, particularly on how the posterior for the utility function is generated. Furthermore, we additionally suggest ways of how the methodology can be turned sequential so as to be made useful for adaptive design situations.
In our future work we expect to provide comparisons of these differing variants of likelihood-free simulation-based optimal designs.

\section{Simulation-based optimal design}

\label{sim_based_optimal_design}

\subsection{Expected utility maximization}

First we consider the case where no prior observations are available. The (future) data $\zb \in \Y$ are sampled according to a known probability model which can be described by the likelihood function $p(\zb|\thetab,\db)$. The likelihood function depends on the parameters $\thetab \in \Theta$ and the chosen design $\db \in \Xi$. We assume that the parameters follow a prior distribution $p(\thetab)$ which does not depend on the design $\db$. The posterior distribution of the parameters is denoted by $p(\thetab|\zb,\db) \propto p(\zb|\thetab,\db) p(\thetab)$, and $p(\zb|\db) = \int_{\thetab \in \Theta} p(\zb|\thetab,\db) p(\thetab) d \thetab$ is the prior predictive distribution.

The general aim of simulation-based optimal design is to find the optimal configuration  $\db^* = \arg \underset{\db}{\sup} \: U(\db)$ for the expected utility integral
\begin{eqnarray}
    U(\db) & = & \int_{\zb \in \Y} \left( \int_{\thetab \in \Theta} u(\zb,\db,\thetab) p (\thetab|\zb,\db) d \thetab \right) p(\zb|\db) d \zb \notag\\
    & = & \int_{\zb \in \Y} \int_{\thetab \in \Theta} u(\zb,\db,\thetab) p (\zb,\thetab|\db) d \thetab d \zb \notag\\
    & = & \int_{\zb \in \Y} \int_{\thetab \in \Theta} u(\zb,\db,\thetab) p (\zb|\thetab,\db) p(\thetab) d \thetab d \zb. \label{expected_utility_integral_1}
\end{eqnarray}
The utility function $u(.)$ may depend on the data $\zb$, the design $\db$, and the parameters $\thetab$. This is the setting considered e\@.g\@. by \citet{mue:sim} and \citet{mue-etal:inh}.

In many cases the integrals are analytically intractable and numerical solution techniques fail due to the high dimension of the variables. However, if it is possible to obtain a sample $\{\zb^{(t)},\thetab^{(t)}, \: t=1,\ldots,T \}$ from $p(\zb,\thetab|\db)$, for example by sampling $\thetab$ from its prior distribution and $\zb|\thetab,\db$ from the probability model, a straightforward way to approximate $U(\db)$ is to use Monte Carlo integration:
\begin{equation*}
    U(\db) \approx \hat{U}(\db) = \frac{1}{T} \sum_{t=1}^T u(\zb^{(t)},\db,\thetab^{(t)}).
\end{equation*}

%Monte Carlo integration does not suffer from the curse of dimensionality as numerical integration techniques do. If the variance of $\zb,\thetab|\db$ is finite, then the variance of $\hat{U}(\db)$ decreases with a rate of $1/T$, regardless of the dimensions of $\zb$ and $\thetab$ (see \citealt{liu:mon}). This is why Monte Carlo integration has such a great appeal, especially for high-dimensional problems.

In a second step, stochastic optimization algorithms may then be performed on $\hat{U}(\db)$ to find the optimum. Possible algorithms include steepest-ascent-type algorithms, global optimization methods such as simulated annealing, genetic algorithms of various kinds, global random search or exchange algorithms, see e\@.g\@. \citet{zhi-zhi:sto}. Depending on the computational effort to obtain the draws $\{\zb^{(t)},\thetab^{(t)}, \: t=1,\ldots,T\}$ and to evaluate the function  $u(\zb^{(t)},\db,\thetab^{(t)})$, this can be time-consuming because every optimization step requires several evaluations of $\hat{U}(\db)$. Specific stochastic optimization algorithms have been developed that minimize the computational costs by reducing the number of evaluations, for example the \emph{simultaneous perturbation stochastic approximation (SPSA)} algorithm proposed by \citet{spa:imp,spa:ove}. For a discussion of these methods in the context of simulation-based design see e\@.g\@. \citet{hua-mar:non}. However, even these improved algorithms may still be slow, depending on the situation, and they may exhibit slow convergence for certain shapes of the utility function.

\subsection{MCMC algorithm}

\citet{mue:sim} proposes an alternative way to tackle this optimization problem that combines the simulation as well as the optimization steps.  To implement a stochastic search the integrand in \eqref{expected_utility_integral_1} is regarded as  proportional to a joint probability distribution of the variables $\zb$, $\db$, and $\thetab$:
\begin{equation*}
h(\zb,\db,\thetab) \propto u(\zb,\db,\thetab) p(\zb|\thetab,\db) p(\thetab) \mu(\db),
\end{equation*}
where $\mu(\db)$ is some (usually uniform) measure on the design region. %It may be viewed as a computational device to randomize $\db$.
 If $u(.)$ is positive and bounded, then $h(.)$ is a proper pdf and
it is  straightforward to use Markov chain Monte Carlo (MCMC) methods such as Metropolis Hastings (MH) to generate draws from this distribution.

Integrating over $\thetab$ and $\zb$, we get
\begin{equation*}
U(\db) \propto \int_{\zb \in \Y} \int_{\thetab \in \Theta} h(\zb,\db,\thetab) d \thetab d \zb,
\end{equation*}
so the marginal distribution of $\db$ is proportional to the expected utility function. Therefore, a strategy to find the optimum design is to sample from $h(\zb,\db,\thetab)$, retain the draws of $\db$, and then search for the mode of the marginal distribution of $\db$ by inspecting the draws.

There are many ways to generate draws from $h(\zb,\db,\thetab)$, for example via Gibbs sampling or by using a hybrid Gibbs/MH sampler. For a review of MCMC sampling schemes see \citet{tie:mar}.

\citet{mue:sim} uses a simple MH scheme with the following proposal distribution for a new draw $\parms'=(\zb',\db',\thetab')$ from
$\parms^{(t-1)}=(\zb^{(t-1)},\db^{(t-1)},\thetab^{(t-1)})$
\begin{equation*}
 q(\parms' | \parms^{(t-1)}) = p(\zb'|\thetab',\db') p(\thetab') g(\db'|\db^{(t-1)}),
\end{equation*}
where $g(\db'|\db^{(t-1)})$ is a random walk proposal for $\db$, and the joint distribution of $\zb$ and $\thetab$ is used as independence proposal for the data and the parameters. Usually it is not hard to sample from this proposal. Specifying the proposal distribution in this way leads to the acceptance probability
\begin{equation*}
\alpha = \min \left( 1, \frac{u(\parms')}{u(\parms^{(t-1)})} \frac{g(\db^{(t-1)}|\db')}{g(\db'|\db^{(t-1))}} \right).
\end{equation*}
Note that the terms $p(\zb'|\thetab',\db') p(\thetab')$ and $p(\zb^{(t-1)}|\thetab^{(t-1)},\db^{(t-1)}) p(\thetab^{(t-1)})$ do not appear in this formula.
%Then a uniform random number $U$ is drawn, and one sets $(\zb^{(t)},\db^{(t)},\thetab^{(t)}) = (\zb',\db',\thetab')$ if $U < \alpha$; otherwise one sets $(\zb^{(t)},\db^{(t)},\thetab^{(t)}) = (\zb^{(t-1)},\db^{(t-1)},\thetab^{(t-1)})$.

However, if $p(\zb,\thetab|\db)$ is rather flat or $u(.)$ is peaked in regions where $p(\zb,\thetab|\db) = p(\zb|\thetab,\db) p(\thetab)$ is relatively low, then this proposal distribution will lead to a poorly performing sampler. Thus, in some cases it might be preferable to use customized proposal densities, which better reflect the shape of the true pdf $h(\zb,\db,\thetab)$. It is also possible to set up a (hybrid) Gibbs scheme with conditional draws of $\zb$, $\db$, and $\thetab$. \citet{jac-etal:max} implement both strategies.

There are several advantages of the MCMC scheme over other standard stochastic optimization approaches: First, as \citet{mue:sim} states, it provides a unified framework for simulation and optimization. Second,  only one draw of $\zb$ and $\thetab$  is sampled at each new design point $\db$, rather than $T$, thereby relieving the computational burden.
%Once the MCMC sampler has found a region of high density, it will stay there for a while due to the Markov chain's autocorrelation. Therefore, the draws automatically concentrate in regions of high expected utility.%
Further \citet{sol-etal:res} note that the output of the MCMC sampler provides a ``map'' of useful designs and not just one single maximum design.

\subsection{Simulated annealing}

\label{simulated_annealing}

A drawback of the basic MCMC algorithm is that it may be very hard to find the mode of the marginal distribution of $\db$ from the MCMC draws, especially in higher-dimensional design spaces and if the expected utility surface is rather flat.

To facilitate the detection of the mode \citet{mue:sim} suggests a method  which is similar to simulated annealing (see e\@.g\@. \citealt{laa-aar:sim}, \citealt{kir-etal:opt}) in that the target distribution is the expected utility function taken to a high power. Then all draws of the MCMC sampler will cluster tightly around the mode, and the mean of the draws will be a good approximation to the mode. In order to achieve that, the target distribution is modified in the following way (augmented model):
\begin{equation}
h_J(\zb_{1:J},\db,\thetab_{1:J}) \propto \prod_{j=1}^J \phi(\zb_j,\db,\thetab_j)  \mu(\db), \label{simulated_annealing_target}
\end{equation}
where
\begin{equation}
\phi(\zb_j,\db,\thetab_j) = u(\zb_j,\db,\thetab_j) p(\zb_j|\thetab_j,\db) p(\thetab_j). \label{target_SMC_1}
\end{equation}

Integrating over $\zb_1,\ldots,\zb_J$ and $\thetab_1,\ldots,\thetab_J$ leads to a marginal distribution proportional to $U^J(\db)$. Therefore, $1/J$ might be regarded as annealing temperature. As $J \rightarrow \infty$ $U^J(\db)$ collapses to a point mass at $\db = \db^{*}$. Thus, for every draw of $\db$ one has to take $J$ independent draws of $\zb$ and $\thetab$ and multiply the utilities.

If there are several modes, there is a danger of getting trapped in one local mode if $J$ is too large. \citet{mue-etal:inh} propose an extension of the basic algorithm where $J$ is gradually increased over time as the algorithm progresses, thereby forming an inhomogeneous Markov chain of the variables. In the beginning the MCMC sampler explores the whole design space so that no local mode is missed. Subsequently the MCMC draws concentrate more and more around one of the highest modes. The acceptance probability formula at time $t$ is
\begin{equation*}
\alpha = \min \left( 1, \frac{\prod_{j=1}^{J(t)} u(\zb_j',\db_j',\thetab_j')}{\prod_{j=1}^{J(t)} u(\zb_j^{(t-1)},\db_j^{(t-1)},\thetab_j^{(t-1)})} \frac{g(\db^{(t-1)}|\db')}{g(\db'|\db^{(t-1))}} \right). %\label{acceptance_probability_simulated_annealing}
\end{equation*}
\citet{mue-etal:inh} show that the inhomogeneous Markov chain produced by this sampler is still strongly ergodic if the ``cooling'' schedule in $J$ is logarithmic. If $J(t) > J(t-1)$, then values $\{(\zb_j^{(t-1)}, \thetab_j^{(t-1)}), \: j  = J(t-1)+1, \ldots, J(t)\}$ are generated by resampling $\{(\zb_j^{(t-1)}, \thetab_j^{(t-1)}), \: j = 1, \ldots, J(t-1)\}$.

%Further improvements have been suggested to explore all modes of the posterior and to avoid to get stuck in local modes, e\@.g\@. in \citet{amz-etal:int} and \citet{rui-etal:evo}. However, all of the techniques presented in this section come with additional computing costs that diminish the appeal of the MCMC algorithm.

\subsection{Particle methods}

\label{particle_methods}

Particle methods are a flexible alternative to MCMC algorithms for sampling from target distributions that change over time. This is also the case for our distribution of interest $h_J(\zb_{1:J},\db,\thetab_{1:J})$ if $J$ is increasing over time. \citet{amz-etal:int} present two interacting particle systems algorithms for the goal of sampling from $h_J(.)$: a classical \emph{sampling importance resampling} algorithm and a computationally more efficient \emph{resampling-Markov} algorithm with weight updates. A different approach is taken by \citet{kue-etal:smc} who employ \emph{sequential Monte Carlo (SMC)} ideas going back to \citet{del-etal:seq}.

\subsubsection{Sampling importance resampling algorithm}

Particle systems usually consist of $N$ particles which are propagated over time. The general sampling importance resampling algorithm is composed of three steps at each time $t$ ($t=1,\ldots,T$).

The \textbf{first step} is a common importance sampling step to obtain an approximate weighted sample from the target distribution. For each particle $n=1,\ldots,N$, at time $t$ a new tuple $\parms_{n,J(t)}^{(t)} = (\zb_{n,1:J(t)}^{(t)},\db_n^{(t)},\thetab_{n,1:J(t)}^{(t)})$ is generated by sampling from the importance function
\begin{equation}
q_{IS}(\parms_{n,J(t)}^{(t)}|\db_n^{(t-1)}) = g_{IS}(\db_n^{(t)}|\db_n^{(t-1)}) \prod_{j=1}^{J(t)} p(\zb_{n,j}^{(t)}|\thetab_{n,j}^{(t)},\db_n^{(t)}) p(\thetab_{n,j}^{(t)}). \label{proposal_SIR_1}
\end{equation}
Therefore, the importance weights are computed as
\begin{equation}
    w_{n}^{(t)} = \frac{h_{J(t)}(\parms_{n,J(t)}^{(t)})}{q_{IS}(\parms_{n,J(t)}^{(t)}|\db_n^{(t-1)})} \propto
        \frac{\prod_{j=1}^{J(t)} u(\zb_{n,j}^{(t)},\db_n^{(t)},\thetab_{n,j}^{(t)})}{g_{IS}(\db_n^{(t)}|\db_n^{(t-1)})}, \quad n=1,\ldots,N. \label{weight_SIR_1}
\end{equation}
After normalizing the weights, i\@.e\@. $W_{n}^{(t)} = w_{n}^{(t)} / \sum_{i=1}^N w_{i}^{(t)}$, a weighted sample is obtained which may serve as a discrete approximation to the target distribution:
\begin{equation*}
h_{J(t)}(d \bm{\vartheta}) \approx \sum_{n=1}^N W_n^{(t)} \delta_{\parms_{n,J(t)}^{(t)}}(d \parms).
\end{equation*}
Here, $\delta_{X}(dx)$ denotes the delta-Dirac measure that concentrates all the mass at point $X$. Moreover, one can use the approximation
\begin{equation*}
\mathbb{E}_{h_{J(t)}}(\varphi) = \int \varphi(\parms) h_{J(t)}(\parms) d \parms \approx \sum_{n=1}^N W_n^{(t)} \varphi(\bm{\vartheta}_{n,J(t)}^{(t)}) = \hat{\mathbb{E}}_{h_{J(t)}}(\varphi)
\end{equation*}
for any measurable and bounded function $\varphi$. The variance of this estimator is approximately
$$ \mathbb{V}_{q_{IS}}(\hat{\mathbb{E}}_{h_{J(t)}}(\varphi)) \approx  \mathbb{V}_{h_{J(t)}} (\varphi) \left( 1 + \mathbb{V}_{q_{IS}} (w_n^{(t)}(\parms_{n,J(t)}^{(t)})) \right) \biggl/ N,$$ see \citet{liu:mon}. If an i\@.i\@.d\@. sample $\{\tilde{\parms}_{1,J(t)},\ldots,\tilde{\parms}_{N,J(t)}\}$ from the target distribution $h_{J(t)}$ were available, the variance of the estimator would be $\displaystyle \mathbb{V}_{h_{J(t)}}(\hat{\mathbb{E}}_{h_{J(t)}}(\varphi)) = \mathbb{V}_{h_{J(t)}} (\varphi) / N$. Hence, the variance of the estimator is directly proportional to the variance of the weights. The quantity
\begin{equation*}
\ess = N \biggl/ \left( 1 + \mathbb{V}_{q_{IS}} (w_n^{(t)}(\bm{\vartheta}_{n,J(t)}^{(t)})) \right)
\end{equation*}
is therefore termed the \emph{effective sample size}. Roughly speaking, it says that $N$ weighted samples drawn from the importance distribution are worth $\ess$ i\@.i\@.d\@. samples drawn directly from the target distribution. The effective sample size can be conveniently estimated by
\begin{equation*}
\widehat{\ess} = 1 \biggl/ \sum_{n=1}^N \left( W_n^{(t)} \right)^2.
\end{equation*}
\citet{amz-etal:int} recommend to use an importance distribution that closely resembles the target distribution, so the variance of $g(\db_n^{(t)}|\db_n^{(t-1)})$ decreases to 0 with rate $1/J$.

The \textbf{second step} is a selection or resampling step. The particles $\{\parms_{n,J(t)}^{(t)}, \: n=1,\ldots,N\}$, are resampled according to a multinomial distribution with weights $W_n^{(t)}$. The new sample $\{\bar{\parms}_{n,J(t)}^{(t)} = (\bar{\zb}_{n,1:J(t)}^{(t)},\bar{\db}_n^{(t)},\bar{\thetab}_{n,1:J(t)}^{(t)}), \: n=1,\ldots,N\}$ is also an approximate sample from the target distribution with equal weights $W_n^{(t)} = 1/N$. On the one hand, the resampling step reduces the \ess, on the other hand, each resampling step is an additional approximation that might deteriorate the approximation quality. Some particles with low weights drop out of the sample after the resampling step, leading to a concentration of particles around few design points.

As a \textbf{third step}, \citet{amz-etal:int} employ a Metropolis-Hastings step to enrich the sample. It would be sufficient to implement only the first (and second) step for the sampler to be valid. However, as mentioned above, this strategy often leads to a strong concentration of design points because many particles are discarded over time. In order to allow the sampler to explore new regions, an MH step is added. \citet{amz-etal:int} suggest to use a proposal distribution for the design points $g_{MH}(\db'_n|\bar{\db}_n^{(t)})$ that has a relatively high variance to make it possible to break out and detect other modes. Using the joint distribution of $\zb$ and $\thetab$ again as independence proposal for these variables (i\@.e\@. $\zb'_{n,1:J(t)},\thetab'_{n,1:J(t)} | \db'_n \sim \prod_{j=1}^{J(t)} p(\zb'_{n,j}|\thetab'_{n,j},\db'_n) p(\thetab'_{n,j})$),  the acceptance probability for the MH step at particle $n$ ($n = 1,\ldots,N$) is given as
\begin{equation*}
\alpha = \min \left( 1, \frac{\prod_{j=1}^{J(t)} u(\zb'_{n,j},\db'_n,\thetab'_{n,j})}{\prod_{j=1}^{J(t)} u(\bar{\zb}_{n,j}^{(t)},\bar{\db}_n^{(t)},\bar{\thetab}_{n,j}^{(t)})} \frac{g(\bar{\db}_n^{(t)}|\db'_n)}{g(\db'_n|\bar{\db}_n^{(t)})} \right).
\end{equation*}
%Set $\db_n^{(t)} = \db'_n$ with probability $\alpha$ and set $\db_n^{(t)} = \bar{\db}_n^{(t)}$ with probability $1 - \alpha$.

\subsubsection{Resampling-Markov algorithm}

\citet{amz-etal:int} also propose a computationally less demanding alternative to the sampling importance resampling algorithm. After initialization the importance sampling step is omitted in all subsequent iterations of the sampler. Assume that at time $t-1$ a sample approximately drawn from $h_{J(t-1)}$ is available. If $J(t) > J(t-1)$ additional values $\{(\zb_{n,j}^{(t-1)},\thetab_{n,j}^{(t-1)}), \: j = J(t-1) + 1,\ldots J(t)\}$ for each particle are drawn from $\displaystyle \zb^{(t-1)}_{n,J(t-1)+1:J(t)},\thetab^{(t-1)}_{n,J(t-1)+1:J(t)} | \db^{(t-1)}_n \sim \prod_{j=J(t-1)+1}^{J(t)} p(\zb^{(t-1)}_{n,j}|\thetab^{(t-1)}_{n,j},\db^{(t-1)}_n) p(\thetab^{(t-1)}_{n,j})$ and the  weights are computed
\begin{equation}
w_n^{(t)} \propto w_n^{(t-1)} \prod_{j=J(t-1)+1}^{J(t)} u(\zb_{n,j}^{(t-1)},\db_n^{(t-1)},\thetab_{n,j}^{(t-1)}). \label{weight_RM_1}
\end{equation}
The resampling and the Metropolis-Hastings steps are performed as before. After each resampling step all the weights are $1/N$, $w_n^{(t-1)} = 1/N$ in the formula above. The weight at iteration $t$ is just proportional to the product of the utilities of the newly sampled values. This algorithm avoids the complete (importance) sampling of all variables at every iteration. However, it may be harder to explore complex utility surfaces with this method.

\subsubsection{Sequential Monte Carlo algorithm}

\label{SMC_algorithm}

\citet{kue-etal:smc} apply the sequential Monte Carlo sampler framework of \citet{del-etal:seq}. An auxiliary augmented target distribution
\begin{equation*}
 \tilde{\pi}_t(\mathbf{x}^{(1:t)}) = \pi_t(\mathbf{x}^{(t)}) \prod_{k=t-1}^1 L_k(\mathbf{x}^{(k+1)},\mathbf{x}^{(k)})
\end{equation*}
is constructed which assumes the distribution $\pi_t(\mathbf{x}^{(t)})$ as its marginal. Importance sampling is performed on the complete sample $\mathbf{x}^{(1:t)}$, even though the actual interest is only in the marginal distribution of $\mathbf{x}^{(t)}$. The kernel function $L_k(.,.)$ is called \emph{backward kernel}. The choice of $L_k(.,.)$ is arbitrary but may affect sampling  efficiency. For a discussion of expedient backward kernels see \citet{del-etal:seq}. The importance distribution of the complete sample $\mathbf{x}^{(1:t)}$ is designed in a similar manner as the augmented target distribution:
\begin{equation*}
q_t(\mathbf{x}^{(1:t)}) = q_1(\mathbf{x}^{(1)}) \prod_{k=2}^t K_k(\mathbf{x}^{(k-1)},\mathbf{x}^{(k)}).
\end{equation*}

For some particle $n \in \{1,\ldots,N\}$, $\mathbf{x}_{n}^{(t)}$ is sampled from the \emph{forward kernel} $K_t(\mathbf{x}_n^{(t-1)},\mathbf{x}_n^{(t)})$ and therefore only depends on $\mathbf{x}_n^{(t-1)}$ and not on the whole particle history. It is possible to calculate the weights at time $t$ recursively:
\begin{equation*}
w_n^{(t)}(\mathbf{x}_n^{(1:t)}) = w_n^{(t-1)}(\mathbf{x}_n^{(1:t-1)}) \frac{\pi_t(\mathbf{x}_n^{(t)})}{\pi_{t-1}(\mathbf{x}_n^{(t-1)})} \frac{L_{t-1}(\mathbf{x}_n^{(t)},\mathbf{x}_n^{(t-1)})}{K_t(\mathbf{x}_n^{(t-1)},\mathbf{x}_n^{(t)})}.
\end{equation*}
If the target densities are only known up to a normalizing constant,  the weights have to be normalized at the end to obtain an approximation to $\tilde{\pi}(\mathbf{x}^{(1:t)})$. Since $\pi(\mathbf{x}^{(t)})$ is just the marginal distribution of $\tilde{\pi}(\mathbf{x}^{(1:t)})$, these weights can also be used to approximate the distribution of $\mathbf{x}^{(t)}$.

SMC algorithms offer a flexible way to slowly ``approach'' a complex target distribution via a sequence of intermediate target distributions. The weights have to be adjusted properly at each time step. The idea is to start with a target distribution $\pi_1$ which is easy to approximate by an importance distribution. If the target distribution changes only gradually from time $t-1$ to time $t$, then slight perturbations of the sample induced by the forward kernel should also be good approximations to the new target distribution. The error of sampling from the wrong distribution is captured by the weights.

The variance of the weights usually increases over time so that the $\ess$ goes down. A common strategy to remedy this problem is to perform a resampling step as soon as the $\ess$ falls below some threshold value (often $N/2$).

\citet{kue-etal:smc} apply the SMC methodology in a particular situation with Shannon information as utility function. In this application    the parameters $\thetab$ can be integrated out of the expected utility integral \eqref{expected_utility_integral_1} which simplifies the problem to some extent. In a more general setting where this integration is not feasible   the following sequence of target distributions is sampled:
\begin{equation*}
h_{J(t),\nu_t}(\zb_{1:J(t)},\db,\thetab_{1:J(t)}) \propto \left( \prod_{j=1}^{J(t)-1} \phi(\zb_j,\db,\thetab_j) \right) \phi(\zb_{J(t)},\db,\thetab_{J(t)})^{\nu_t} \mu(\db).
\end{equation*}

The monotonically increasing inverse annealing temperature $J(t)$ is integer-valued, and $J(t) \leq J(t-1)+1$, so $J$ increases by at most 1 at each time step. In order to allow for more finely graduated non-integer jumps of the inverse annealing temperature, \citet{kue-etal:smc} introduce the variable $\nu_t \in [0,1]$. The original target density \eqref{simulated_annealing_target} is obtained if $\nu_t=1$. Whenever $\nu_t$ reaches 1, the dimension of the target distribution increases, $J(t) = J(t-1)+1$, and $\nu_t$ is set back to 0 again. In all other cases $J(t) = J(t-1)$.

The forward kernel suggested by \citet{kue-etal:smc}  is
\begin{align*}
&K_t \left( \parms_{n,J(t-1)}^{(t-1)}, \parms_{n,J(t)}^{(t)} \right) = \quad \mathcal{K}_{J(t)-1,1} \left(  \parms_{n,J(t)-1}^{(t-1)},  \parms_{n,J(t)-1}^{(t)} \right) q_{\nu_t} \left(\zb_{n,J(t)}^{(t)},\thetab_{n,J(t)}^{(t)} | \db_n^{(t)} \right),
\end{align*}
where  $\parms_{n,j}^{(t)}=( \zb_{n,1:j}^{(t)},\db_{n}^{(t)},\thetab_{n,1:j}^{(t)})$,  $\mathcal{K}_{J(t)-1,1}$ is the Metropolis-Hastings kernel with invariant distribution $h_{J(t)-1,1}$, and $q_{\nu_t} \left(\zb_{n,J(t)}^{(t)},\thetab_{n,J(t)}^{(t)} | \db_n^{(t)}\right)$ is an independence proposal for $\zb_{n,J(t)}^{(t)}$ and $\thetab_{n,J(t)}^{(t)}$.

Selecting the backward kernel as
\begin{align*}
&L_{t-1} \left( \parms_{n,J(t)}^{(t)}, \parms_{n,J(t-1)}^{(t-1)} \right) = \mathcal{K}_{n,J(t)-1,1} \left( \parms_{n,J(t)-1}^{(t-1)}, \parms_{n,J(t)-1}^{(t)} \right) \cdot  \frac{h_{J(t-1),\nu_{t-1}} \left(\parms_{n,J(t-1)}^{(t-1)}\right)}{h_{J(t)-1,1}\left(\parms_{n,J(t)-1}^{(t)}\right)}
\end{align*}
leads to the incremental weights
\begin{equation}
w_n^{(t)} \propto w_n^{(t-1)} \frac{\phi\left(\zb_{n,J(t)}^{(t)},\db^{(t)},\theta_{n,J(t)}^{(t)}\right)^{\nu_t}}{q_{\nu_t} \left(\zb_{n,J(t)}^{(t)},\db^{(t)},\theta_{n,J(t)}^{(t)}\right)}. \label{weight_SMC_1}
\end{equation}

A slowly evolving target distribution usually implies a better fitting importance distribution, so the $\ess$ deteriorates less rapidly and there have to be fewer resampling steps.

\citet{kue-etal:smc} use $q_{\nu_t} \left(\zb_{n,J(t)}^{(t)},\thetab_{n,J(t)}^{(t)} | \db_n^{(t)}\right) = p \left(\zb_{n,J(t)}^{(t)},\thetab_{n,J(t)}^{(t)} | \db_n^{(t)}\right)$. They mention that this choice of proposal distribution seems to work well when $\nu_t=1$. If $\nu_t < 1$, the importance weights tend to exhibit a high variance. Furthermore, $\zb_{n,J(t)}^{(t)}$ and $\thetab_{n,J(t)}^{(t)}$ are sampled independently from the previous sample at all time steps. Ideally, they should only be sampled independently when $J(t)$ makes a jump, i\@.e\@. when $J(t) = J(t-1)+1$. \citet{kue-etal:smc} therefore propose another sampling scheme in which they use different forward and backward kernels depending on whether $J(t) = J(t-1) + 1$ or $J(t) = J(t-1)$.

\section{Simulation-based optimal design with prior observations}

The algorithms presented in Section \ref{sim_based_optimal_design} have to be modified slightly if prior observations $\yos = \{\mathbf{y}_i, \, i=1,\ldots,s\}$ measured at the design points $\xios = \{\bm{\xi}_i, \, i=1,\ldots,s\}$ are available for $s$ measurements collected in the past (cf\@. \citealt{mue-etal:inh}). If we assume that
\begin{equation*}
p (\thetab|\{\zb,\yos\},\{\db,\xios\}) \propto p(\zb|\thetab,\db) \left( \prod_{i=1}^s p(\mathbf{y}_i|\thetab,\bm{\xi}_i) \right) p(\thetab) = p(\zb|\thetab,\db) p(\yos|\thetab,\xios) p(\thetab),
\end{equation*}
then the expected utility integral \eqref{expected_utility_integral_1} changes to
\begin{eqnarray*}
    U(\db) & = & \int_{\zb \in \Y} \left( \int_{\thetab \in \Theta} u(\{\zb,\yos\},\{\db,\xios\},\thetab) p (\thetab|\{\zb,\yos\},\{\db,\xios\}) d \thetab \right) p(\zb|\yos,\{\db,\xios\}) d \zb \notag\\
    & = & \int_{\zb \in \Y} \left( \int_{\thetab \in \Theta} u(\{\zb,\yos\},\{\db,\xios\},\thetab) \frac{p(\zb|\thetab,\db) \left( \prod_{i=1}^s p(\mathbf{y}_i|\thetab,\bm{\xi}_i) \right) p(\thetab)}{p(\{\zb,\yos\}|\{\db,\xios\})} d \thetab \right) \times \\
    & & \qquad \quad p(\zb|\yos,\{\db,\xios\}) d \zb \notag\\
    & = & \int_{\zb \in \Y} \left( \int_{\thetab \in \Theta} u(\{\zb,\yos\},\{\db,\xios\},\thetab) \frac{p(\zb|\thetab,\db)}{p(\zb|\yos,\{\db,\xios\})} \frac{\left( \prod_{i=1}^s p(\mathbf{y}_i|\thetab,\bm{\xi}_i) \right) p(\thetab)}{ p(\yos|\xios)} d \thetab \right) \times \\
    & & \qquad \quad p(\zb|\yos,\{\db,\xios\}) d \zb \notag\\
    & = & \int_{\zb \in \Y} \int_{\thetab \in \Theta} u(\{\zb,\yos\},\{\db,\xios\},\thetab) p (\zb|\thetab,\db) \, p(\thetab|\yos,\xios) d \thetab d \zb. \label{expected_utility_integral_2}
\end{eqnarray*}

If it is possible to sample from the posterior distribution $p(\thetab|\yos,\xios)$, then all the algorithms introduced in the previous section can still be used. The only difference is that instead of $p(\thetab)$ one has to use $p(\thetab|\yos,\xios)$ for  sampling  the parameters. However,   MCMC and importance sampling algorithms are usually employed when  a closed-form representation of the posterior distribution cannot be obtained. Hence, the algorithms presented in Section \ref{sim_based_optimal_design} have to be generalized. Instead of
\begin{equation*}
q(\zb'_{1:J},\db',\thetab'_{1:J}|\db^{(t-1)}) = \prod_{j=1}^J p(\zb'_j|\thetab'_j,\db') p(\thetab'_j|\yos,\xios) g(\db'|\db^{(t-1)})
\end{equation*}
the more general proposal distribution
\begin{equation*}
q_k(\zb'_{1:J},\db',\thetab'_{1:J}|\db^{(t-1)}) = \prod_{j=1}^J p(\zb'_j|\thetab'_j,\db') k(\thetab'_j|\yos,\xios) g(\db'|\db^{(t-1)}).
\end{equation*}
is used.
Thus $\thetab'_j$ is not sampled from the correct posterior,  but from an approximate proposal distribution which should resemble the posterior distribution as closely as possible. Common choices for these proposals are normal or t random walk or independence proposals, where the scale is proportional to the inverse of the Hessian of the likelihood or the unnormalized posterior.

The target distribution changes from \eqref{simulated_annealing_target} to
\begin{eqnarray}
& h_J(\zb_{1:J},\db,\thetab_{1:J}) \propto  \prod_{j=1}^J u(\{\zb_j,\yos\},\{\db,\xios\},\thetab_j) p(\zb_j|\thetab_j,\db) p(\thetab_j|\yos,\xios) \mu(\db) \propto \notag \\
 & \phantom{abc} \propto  \prod_{j=1}^J u(\{\zb_j,\yos\},\{\db,\xios\},\thetab_j) p(\zb_j|\thetab_j,\db) p(\yos|\thetab_j,\xios) p(\thetab_j) \mu(\db). \label{target_adaptive}
\end{eqnarray}
In a standard MCMC annealing algorithm this would lead to the acceptance probability (cf\@. \citealt{mue-etal:inh})
\begin{align}
\alpha = \min \biggl( 1, & \frac{\prod_{j=1}^{J(t)} u(\{\zb_j',\yos\},\{\db',\xios\},\thetab_j')}{\prod_{j=1}^{J(t)} u(\{\zb_j^{(t-1)},\yos\},\{\db^{(t-1)},\xios\},\thetab_j^{(t-1)})} \frac{\prod_{j=1}^{J(t)} p(\yos|\thetab'_j,\xios) p(\thetab'_j)}{\prod_{j=1}^{J(t)} p(\yos|\thetab^{(t-1)}_j,\xios) p(\thetab^{(t-1)}_j)} \times \notag\\
 & \frac{\prod_{j=1}^{J(t)} k(\thetab^{(t-1)}_j|\yos,\xios)}{\prod_{j=1}^{J(t)} k(\thetab'_j|\yos,\xios)} \frac{g(\db^{(t-1)}|\db')}{g(\db'|\db^{(t-1))}} \biggr). \label{acceptance_probability_adaptive}
\end{align}

Similar modifications are necessary for applying the  sampling importance resampling algorithm. The importance distribution \eqref{proposal_SIR_1} could  be changed to
\begin{equation*}
q_{IS,k}(\parms_{n,J(t)}^{(t)}|\db_n^{(t-1)}) = g_{IS}(\db_n^{(t)}|\db_n^{(t-1)}) \prod_{j=1}^{J(t)} p(\zb_{n,j}^{(t)}|\thetab_{n,j}^{(t)},\db_n^{(t)}) k(\thetab_{n,j}^{(t)}|\yos,\xios),
\end{equation*}
implying a change of the weights from \eqref{weight_SIR_1} to
\begin{equation}
 w_{n}^{(t)} = \frac{h_{J(t)}(\parms_{n,J(t)}^{(t)})}{q_{IS}(\bm{\vartheta}_{n,J(t)}^{(t)}|\db_n^{(t-1)})} \propto
        \frac{\prod_{j=1}^{J(t)} u(\{\zb_{n,j}^{(t)},\yos\},\{\db_n^{(t)},\xios\},\thetab_{n,j}^{(t)}) p(\yos|\thetab_{n,j}^{(t)},\xios) p(\thetab_{n,j}^{(t)})}{g_{IS}(\db_n^{(t)}|\db_n^{(t-1)}) \prod_{j=1}^{J(t)} k(\thetab_{n,j}^{(t)}|\yos,\xios)}. \label{weight_SIR_adaptive}
\end{equation}
The acceptance probability for the MH step corresponds to formula \eqref{acceptance_probability_adaptive}.

Replacing $p(\thetab_{n,j}^{(t-1)})$ by $k(\thetab_{n,j}^{(t-1)}|\yos,\xios)$ in the proposal distribution of the resampling-Markov algorithm, the weights \eqref{weight_RM_1} change to
\begin{equation}
w_n^{(t)} \propto w_n^{(t-1)} \prod_{j=J(t-1)+1}^{J(t)} \frac{u(\{\zb_{n,j}^{(t-1)},\yos\},\{\db_n^{(t-1)},\xios\},\thetab_{n,j}^{(t-1)}) p(\yos|\thetab_{n,j}^{(t-1)},\xios) p(\thetab_{n,j}^{(t-1)})}{k(\thetab_{n,j}^{(t-1)}|\yos,\xios)}. \label{weight_RM_adaptive}
\end{equation}

The SMC algorithm in Section \ref{SMC_algorithm} also needs only minor adjustments if prior observations are included. The function $\phi(\zb_j,\db,\thetab_j)$ defined by \eqref{target_SMC_1} is modified to
\begin{equation*}
\phi(\zb_j,\db,\thetab_j) = u(\{\zb_j,\yos\},\{\db,\xios\},\thetab_j) p(\zb_j|\thetab_j,\db) p(\yos|\thetab_j,\xios) p(\thetab_j),
\end{equation*}
and the MH kernels $\mathcal{K}_{J(t)-1,1}$ and independence proposals $q_{\nu_t}$ have to be adapted accordingly.

\section{Approximate Bayesian computing (ABC)}

\label{ABC}

\subsection{General methodology}

In complex models where an explicit form of the likelihood function $p(\zb|\thetab,\db) p(\yos|\thetab,\xios)$ is not available or it is very cumbersome to evaluate, one may have to resort to \emph{likelihood-free (LF)} methods, also called \emph{approximate Bayesian computing (ABC)}. These methods can be applied if simulating the data from the probability model is feasible for every parameter $\thetab$. Instances where likelihood-free methods have been successfully applied include biogenetics (\citealt{mar-etal:mar}, \citealt{bea-etal:app}), ecology (\citealt{jab-cha:inf}), quantile distributions (\citealt{dro-pet:lik}), models for extremes (\citealt{erh-smi:app}, \citealt{bor-etal:inf}), and many other cases. For some further examples see \citet{sis-fan:lik}.

The easiest case is likelihood-free rejection sampling. Sampling  from the posterior distribution $p(\thetab|\yb)$ can be accomplished  by drawing the parameters from the prior distribution, $\thetab' \sim p(\thetab)$, drawing a variable $\xb$ from the probability model, $\xb' \sim p(\xb|\thetab')$, and accepting $\thetab'$ if $\xb' \approx \yb$. Thus, direct sampling from the posterior distribution is replaced by sampling from the prior distribution. The efficiency of this approach crucially depends on the similarity between the posterior and the prior distribution, i\@.e\@. the information gain of the posterior compared to the prior distribution.

Likelihood-free MCMC builds upon the same general idea. The original posterior target distribution $p(\thetab|\yb)$ is approximated by the marginal  of the augmented distribution
\begin{equation}
p_{LF} (\thetab,\xb|\yb) \propto p_{\epsilon}(\yb|\xb,\thetab) p(\xb|\thetab) p(\thetab). \label{LF_augmented_posterior}
\end{equation}
 The variable $\xb$, which is sampled together with $\thetab$, is added to the posterior arguments. Integrating over $\xb$ leads to the original posterior distribution if $p_{\epsilon}(\yb|\xb,\thetab)$ is a point mass at the point $\xb=\yb$. Since this event has a very small probability for higher-dimensional discrete distributions and probability zero in the case of continuous distributions, a compromise has to be found between exactness and practicality by adjusting the ``narrowness'' of $p_{\epsilon}(\yb|\xb,\thetab)$. The function $p_{\epsilon}(\yb|\xb,\thetab)$ is usually assumed to be a smoothing kernel density function: $p_{\epsilon}(\yb|\xb,\thetab) = (1/\epsilon) K( (|\Tb(\xb) - \Tb(\yb)|)/\epsilon )$, where $\Tb(.)$ is some low-dimensional statistic of $\yb$ and $\xb$, respectively. The parameter $\epsilon$ controls the tightness of $p_{\epsilon}(\yb|\xb,\thetab)$.
  If $\Tb$ is a sufficient statistic for the parameters of the probability model, integrating  over $\Tb(\xb)$ yields the same distribution   as  integrating out $\xb$.

Popular choices for $K(.)$ are uniform, Gaussian, or Epanechnikov kernels. The uniform kernel is defined by
\begin{equation*}
p_{\epsilon}(\yb|\xb,\thetab) \propto \mathbb{I}_{A_{\epsilon,\yb}}(\xb),
\end{equation*}
where $\mathbb{I}_{A_{\epsilon,\yb}}(\xb)$ is the indicator function which takes the value 1 if $\xb \in A_{\epsilon,\yb}$ and 0 otherwise. The neighborhood set $A_{\epsilon,\yb}$ is defined as $A_{\epsilon,\yb} = \{\xb \in \mathcal{Y}: \; \rho(\Tb(\xb),\Tb(\yb)) \leq \epsilon \}$, where $\rho$ denotes a distance measure between $\Tb(\xb)$ and $\Tb(\yb)$.

The reason for complicating the model artificially is that for the proposal distribution
\begin{equation*}
q_{LF}(\thetab',\xb'|\thetab^{(t-1)},\xb^{(t-1)}) = q_{\thetab}(\thetab'|\thetab^{(t-1)}) p(\xb'|\thetab'),
\end{equation*}
 the likelihood terms $p(\xb^{(t-1)}|\thetab^{(t-1)})$ and $p(\xb'|\thetab')$ cancel out in the acceptance probability
\begin{eqnarray*}
\alpha & = & \min \left( 1, \frac{p_{\epsilon}(\yb'|\xb',\thetab') p(\xb'|\thetab') p(\thetab')}{p_{\epsilon}(\yb^{(t-1)}|\xb^{(t-1)},\thetab^{(t-1)}) p(\xb^{(t-1)}|\thetab^{(t-1)}) p(\thetab^{(t-1)})} \frac{q_{\thetab}(\thetab^{(t-1)}|\thetab') p(\xb^{(t-1)}|\thetab^{(t-1)})}{q_{\thetab}(\thetab'|\thetab^{(t-1)}) p(\xb'|\thetab')} \right) \\
& = & \min \left( 1, \frac{p_{\epsilon}(\yb'|\xb',\thetab') p(\thetab')}{p_{\epsilon}(\yb^{(t-1)}|\xb^{(t-1)},\thetab^{(t-1)}) p(\thetab^{(t-1)})} \frac{q_{\thetab}(\thetab^{(t-1)}|\thetab')}{q_{\thetab}(\thetab'|\thetab^{(t-1)})} \right).
\end{eqnarray*}
and it is no longer necessary to evaluate the likelihood. As in the simple rejection sampling case, if prior and posterior distribution have very different shapes, the rejection rates will be very close to 1.

In order to reduce the variance of the MH acceptance ratio and thereby improve the sampler's performance, \citet{del-etal:ada} propose an augmented likelihood-free posterior distribution given by
\begin{equation*}
p_{LF} (\thetab,\xoM|\yb) \propto \left[ \frac{1}{M} \sum_{m=1}^M p_{\epsilon}(\yb|\xb_m,\thetab) \right] \left[ \prod_{m=1}^M p(\xb_m|\thetab) \right] p(\thetab),
\end{equation*}
see also \citet{sis-fan:lik}. If $M \rightarrow \infty$, then $\frac{1}{M} \sum_{m=1}^M p_{\epsilon}(\yb|\xb_m,\thetab) \rightarrow \int p_{\epsilon}(\yb|\xb,\thetab) p(\xb|\thetab) d \xb$, and the sampler is similar to the ``marginal'' sampler of the marginal of the likelihood-free posterior \eqref{LF_augmented_posterior} where $\xb$ is integrated out.

Choosing a proposal distribution of the form
\begin{equation*}
q(\thetab',\xoM'|\thetab^{(t-1)},\xoM^{(t-1)}) = q_{\thetab}(\thetab'|\thetab^{(t-1)}) \prod_{m=1}^M p(\xb_m'|\thetab')
\end{equation*}
gives the MH acceptance probability
\begin{equation*}
\alpha = \min \left( 1, \frac{\sum_{m=1}^M p_{\epsilon}(\yb'|\xb'_m,\thetab')}{\sum_{m=1}^M p_{\epsilon}(\yb^{(t-1)}|\xb_m^{(t-1)},\thetab^{(t-1)})} \frac{p(\thetab')}{p(\thetab^{(t-1)})} \frac{q_{\thetab}(\thetab^{(t-1)}|\thetab')}{q_{\thetab}(\thetab'|\thetab^{(t-1)})} \right).
\end{equation*}
Note that this algorithm reduces to the original ABC algorithm if $M=1$.
%idea originally from: Andrieu et al. (2008): The expected auxiliary variable method for Monte Carlo simulation

A comprehensive account of likelihood-free MCMC is given in \citet{sis-fan:lik}.

\subsection{ABC for simulation-based optimal design}

\label{ABC_simulation_based_design}

There are two possible ways to deal with non-available likelihood functions in simulation-based optimal design problems when prior observations are present. If these prior observations are accounted for in the target distribution, the likelihood functions do not cancel out in the MH acceptance probability formula (see \eqref{acceptance_probability_adaptive}) or in the weights or weight updates of (sequential) importance samplers (see \eqref{weight_SIR_adaptive} and \eqref{weight_RM_adaptive}).

\subsubsection{Modifying the target distribution}

One solution would be to augment the target distribution \eqref{target_adaptive} to
\begin{eqnarray*}
h_{LF,J}(\parms_j,\mathbf{x}_{1:s,1:M,1:J}) & \propto & \prod_{j=1}^J u(\{\zb_j,\yos\},\{\db,\xios\},\thetab_j) p(\zb_j|\thetab_j,\db) \times \\
 & & \left[ \frac{1}{M} \sum_{m=1}^M p_{\epsilon}(\yos|\xb_{1:s,m,j},\thetab_j) \right] \left[ \prod_{m=1}^M p(\xb_{1:s,m,j}|\thetab_j,\xios) \right] p(\thetab_j) \mu(\db),
\end{eqnarray*}
where $\mathbf{x}_{1:s,1:M}$ denotes a collection of $M$ draws of $\mathbf{x}_{1:s}$.

A sensible and convenient choice for the proposal distribution is
\begin{equation}
q_{LF}(\parms_j',\mathbf{x}_{1:s,1:M,1:J}'|\db^{(t-1)}) = g(\db'|\db^{(t-1)}) \prod_{j=1}^J p(\zb'_j|\thetab'_j,\db') \left[ \prod_{m=1}^M p(\xb_{1:s,m,j}'|\thetab_j',\xios) \right] p(\thetab_j'), \label{proposal_ABC}
\end{equation}
which leads to the MH acceptance probability
\begin{align*}
\alpha = \min \biggl( 1, & \frac{\prod_{j=1}^{J(t)} u(\{\zb_j',\yos\},\{\db',\xios\},\thetab_j')}{\prod_{j=1}^{J(t)} u(\{\zb_j^{(t-1)},\yos\},\{\db^{(t-1)},\xios\},\thetab_j^{(t-1)})} \times \notag\\
 & \frac{\prod_{j=1}^{J(t)} \left[\frac{1}{M} \sum_{m=1}^{M} p_{\epsilon}(\yos|\xb_{1:s,m,j}',\thetab_j') \right]}{\prod_{j=1}^{J(t)} \left[\frac{1}{M} \sum_{m=1}^{M} p_{\epsilon}(\yos|\xb_{1:s,m,j}^{(t-1)},\thetab_j^{(t-1)}) \right]} \frac{g(\db^{(t-1)}|\db')}{g(\db'|\db^{(t-1))}} \biggr).
\end{align*}
Using an importance distribution similar to \eqref{proposal_ABC} results in weights of
\begin{equation*}
 w_{n}^{(t)} \propto
        \frac{\prod_{j=1}^{J(t)} u(\{\zb_{j}^{(t)},\yos\},\{\db^{(t)},\xios\},\thetab_{j}^{(t)}) \left[ \frac{1}{M} \sum_{m=1}^M p_{\epsilon}(\yos|\xb_{1:s,m,j}^{(t)},\thetab_j^{(t)})  \right]}{g_{IS}(\db^{(t)}|\db^{(t-1)})}
\end{equation*}
for the sampling importance resampling sampler (cf\@. equation \eqref{weight_SIR_adaptive}) and
\begin{equation*}
w_n^{(t)} \propto w_n^{(t-1)} \prod_{j=J(t-1)+1}^{J(t)} u(\{\zb_{j}^{(t-1)},\yos\},\{\db^{(t-1)},\xios\},\thetab_{j}^{(t-1)}) \left[ \frac{1}{M} \sum_{m=1}^M p_{\epsilon}(\yos|\xb_{1:s,m,j}^{(t-1)},\thetab_j^{(t-1)}) \right]
\end{equation*}
for the resampling-Markov algorithm (cf\@. equation \eqref{weight_RM_adaptive}). The subscripts $n$ for the $n$\textsuperscript{th} particle are omitted to simplify notation.

If $M$ is low, the acceptance probability will be close to 0 for most iterations so that the Markov chain is highly persistent. If an importance sampler is applied instead, a small $M$ will lead to very imbalanced weights (many weights will be 0) and therefore to a low $\ess$. On the other hand, choosing a sufficiently high $M$ may be computationally prohibitive, in particular if sampling the data is costly.

When using particle methods, rejection control techniques can be employed to alleviate the problem of imbalanced weights (see e\@.g\@. \citealt{liu:mon}, \citealt{sis-etal:seq}, and \citealt{bea-etal:ada}). Particles with low weights are discarded and resampled from the particle set at time $t-1$ according to the weight distribution. Therefore, rejection control techniques may also require considerable additional computational resources if the size of the particle set is required to stay constant over time. Moreover, it may be very hard to select a proper value for $\epsilon$ or a proper schedule $\{\epsilon_t, \; t=1,\ldots,T\}$ (typically monotonically decreasing) and assess its efficiency for a complex target distribution such as $h_{LF,J(t),\epsilon_t}$. There are many other factors that crucially affect the efficiency of the sampler, for example the choice of $M$ or the schedule for $J(t)$.

\subsubsection{Estimating the posterior distribution separately}

A second potential approach to solving the simulation-based design problem with prior observations and non-available likelihood functions is to divide the problem into smaller tasks.

In a first step, the posterior distribution
\begin{equation*}
p(\thetab|\yos,\xios) \propto \prod_{i=1}^s p(\yb_i|\thetab,\bm{\xi}_i) p(\thetab)
\end{equation*}
may be approximated by any suitable method. One possible method is the sequential Monte Carlo algorithm dealing with general ABC problems developed in \citet{del-etal:ada}. Their algorithm automatically adapts the schedule for $\epsilon_t$. The complexity of this algorithm is $O(R)$, where $R$ denotes the number of particles used for approximating $p(\thetab|\yos,\xios)$. An alternative importance sampler for the same task which is based on population Monte Carlo ideas (\citealt{cap-etal:pop}) is proposed by \citet{sis-etal:seq} and \citet{bea-etal:ada}. The latter improve the method of \citet{sis-etal:seq} by taking care of the bias that is engendered by the originally proposed method. Both algorithms also make use of rejection control, whereas \citet{del-etal:ada} use a sufficiently high simulation number $M$ to improve  sampling quality. The algorithm of \citet{bea-etal:ada} has a complexity of $O(R^2)$.

The final particle set delivered by any of these particle methods can be regarded as a discrete approximation to the true posterior distribution. The weights correspond to the probability mass function of the posterior approximation. One may then replace the exact posterior distribution $p(\thetab|\yos,\xios)$ with this approximate distribution $\hat{p}_{LF}(\thetab|\yos,\xios)$, which is the marginal distribution of $\hat{p}_{LF}(\thetab,\mathbf{x}_{1:s,1:M}|\yos,\xios)$, in the algorithms described in Sections \ref{simulated_annealing} and \ref{particle_methods}. Sampling from $\hat{p}_{LF}(\thetab|\yos,\xios)$ is easily accomplished, for example, by discrete sampling from the weighted particle set $\{\thetab_{r}^{(L)},\; r = 1,\ldots,R\}$. The particle sampler for estimating $p(\thetab|\yos,\xios)$ comprises $R$ particles and iterates over $L$ time steps.

A  way to obtain an approximate closed formula for the posterior distribution is the non-parametric conditional density estimation approach of \citet{bon-etal:bay}. They fit a mixture of multivariate normals to the joint distribution of the summary statistics and the parameters $(\Tb(\mathbf{x}),\thetab)$ based on the samples $\{ (\Tb(\mathbf{x}^{(l)}),\thetab^{(l)}), \; l = 1,\ldots,L \}$ and  condition on the summary statistics to get a normal mixture approximation of the posterior. A similar approach is taken by \citet{fan-etal:reg} to obtain an analytic representation of the likelihood function. The likelihood function can then be combined with different priors, for example in order to conduct a sensitivity analysis. Thus, estimating the likelihood instead of the posterior offers greater flexibility. \citet{fan-etal:reg} use the samples $\{ (\Tb(\mathbf{x}^{(l)},\thetab^{(l)}), \; l = 1,\ldots,L \}$ to fit univariate mixtures of normals to the summary statistics. The normal mixtures depend on the parameters $\thetab$. In order to account for the dependence structure between the summary statistics and the parameters, they use  a mixture of Gaussian copulas for the standardized summary statistics and the parameters. Their method works quite well even for higher-dimensional summary statistics.

%If a closed-form (approximate)  for the posterior distribution is available,  again all the methods in Sections \ref{simulated_annealing} and %\ref{particle_methods} can be applied, even the SMC algorithm in Section \ref{SMC_algorithm}, where the prior density function features in the weight %update \eqref{weight_SMC_1}.

\section{Utility function}

\label{utility_function}

The choice of the utility function $u(\{\zb,\yos\},\{\db,\xios\},\thetab)$ is highly problem-specific. The experimenter has great freedom in selecting an appropriate utility function. However, the utility function has to be bounded and  non-negative, otherwise $h_J$ will not be a proper probability distribution.

It is possible to construct very customized utility functions  as \citet{mue-etal:inh} do for their network design examples (a rainfall network and an ozone monitoring network). Their utility functions incorporate information about the number of future observations that are correctly covered by a small confidence interval around the predicted response value for stations that are not in the network, or the number of future observations that are correctly estimated to be above a certain threshold value if interest is in predicting extreme events. Furthermore, they incorporate the sum of squared distances of the future observations from the predicted response values for all stations that are not in the network, and  include the variation of the response surface around its estimated value. Lastly, they also allow for the costs of the number of stations to enter into the utility function.

If the different designs can be interpreted as alternative models,  it is even possible to perform some kind of model  evaluation  and selection   by comparing the simulated $\zb$ to the past observations $\yos$.

There is often a need to base the utility on parameters of the updated posterior distribution $p(\thetab|\{\zb,\yos\},\{\db,\xios\})$, like the variance of $\thetab$ or the Kullback-Leibler divergence. Sometimes it may even be of interest to include parameters of the updated posterior predictive distribution $\int p(\zb_{p}|\thetab,\db_{p}) p(\thetab|\{\zb,\yos\},\{\db,\xios\}) d \thetab$, like the estimated mean response at some points in the design space. These parameters cannot be used in the utility function unless there is an analytical expression for them which is very unusual for the kinds of problems where ABC methods are used. However, if an approximate sample from the posterior distribution $p(\thetab|\yos,\xios)$ is available, for example by performing a particle sampler beforehand, the posterior distribution can be updated rather quickly.

Assume that a weighted set of particles $\{(\thetab_{r}^{(L)}, W_{r}^{(L)}), \; r=1,\ldots,R\}$ distributed according to the approximate distribution $\hat{p}_{LF}(\thetab|\yos,\xios) \approx p(\thetab|\yos,\xios)$ is available.  The %target distribution of the
updated posterior distribution is then

\begin{equation*}
\hat{p}_{LF}(\thetab,\mathbf{x}_{1:M}|\{\zb,\yos\},\{\db,\xios\}) \propto \left[ \frac{1}{M} \sum_{m=1}^M p_{\epsilon}(\zb|\xb_m,\thetab) \right] \left[ \prod_{m=1}^M p(\xb_m|\thetab,\db) \right] \hat{p}_{LF}(\thetab|\yos,\xios).
\end{equation*}
The importance distribution is constructed by sampling $\thetab_{r}^{(L+1)}$ from the approximate posterior and sampling $\{ \xb_{m}^{(L+1)}, \; m = 1,\ldots,M \}$ from $p(\xb_m|\thetab,\db)$, that is
\begin{equation*}
q_{\thetab,LF}(\thetab,\mathbf{x}_{1:M}|\{\zb,\yos\},\{\db,\xios\})  = \left[ \prod_{m=1}^M p(\xb_m|\thetab,\db) \right] \hat{p}_{LF}(\thetab|\yos,\xios),
\end{equation*}
so one arrives at the new importance weights for the updated approximate posterior:
\begin{equation*}
w_r^{(L+1)} \propto \sum_{m=1}^M p_{\epsilon}(\zb|\xb_m,\thetab).
\end{equation*}
This whole procedure has to be repeated $J(t)$ times at each time step $t = 1,\ldots,T$ if a simulated-annealing-type extension is employed. In addition, rejection control techniques may be applied to ensure that only particles with non-zero weights are sampled (see \citealt{liu:mon}).

Given the particles of the updated approximate posterior, one can either construct summary statistics like the variance of $\thetab|\{\zb,\yos\},\{\db,\xios\}$ (cf\@. the Bayesian D-posterior precision utility used by \citealt{dro-etal:seq}) or simulate from the updated posterior predictive distribution $\int p(\zb_{p}|\thetab,\db_{p}) p(\thetab|\{\zb,\yos\},\{\db,\xios\}) d \thetab$ by sampling $\thetab_p^{(1:L)}$ from $\hat{p}_{LF}(\thetab|\{\zb,\yos\},\{\db,\xios\})$ and then sampling $\zb_p^{(1:L)}$ from the likelihood. A simple parameter like the mean response at design point $\db_p$ can then be calculated by averaging the draws of $\zb_p^{(1:L)}$. A single forecast is created by setting $L=1$. \citet{sol-etal:res} follow a similar approach in order to estimate a response variance criterion for their utility function.

An alternative way to update the posterior distribution would be to approximate the likelihood function by a multivariate mixture of normals as in \citet{fan-etal:reg}. It is then straightforward to combine the analytic expressions for the likelihood of $\{\zb,\yos\}$ and for the prior distribution of the parameters to obtain an update of the posterior distribution, for example by creating a sample from the updated posterior by standard MCMC sampling. Calculating the updated posterior distribution, however, may be rather computer intensive, which diminishes the usefulness of this method if a new updated posterior has to be computed for every evaluation of $u(.)$.

A very important standard utility criterion which depends on the updated posterior is the Kullback-Leibler divergence, also often called Shannon information:
\begin{equation*}
\text{KLD}(\zb,\db) = \int_{\thetab \in \Theta} \log \left( \frac{p(\thetab|\{\zb,\yos\},\{\db,\xios\})}{p(\thetab)} \right) p(\thetab|\{\zb,\yos\},\{\db,\xios\}) d \thetab.
\end{equation*}
The KLD may also be computed as distance between $p(\thetab|\{\zb,\yos\},\{\db,\xios\})$ and $p(\thetab|\yos,\xios)$.

\citet{cha-ver:rev} point out that for a standard linear model the average Kullback-Leibler distance $\int_{\zb \in \Y} \text{KLD}(\zb,\db) p(\zb|\db) d \zb$ corresponds to the classical D-criterion. In the case of non-linear models, the average Kullback-Leibler distance still gives a reasonable approximation to the Bayesian D-criterion involving the Fisher information matrix if the posterior distribution of $\thetab$ is approximately normally distributed. Moreover, the Kullback-Leibler distance is more robust than utilities which only depend on the variance of $\thetab|\{\zb,\yos\},\{\db,\xios\}$, especially if the posterior is multi-modal.

Computing the KLD makes it necessary to obtain $p(\thetab|\{\zb,\yos\},\{\db,\xios\})$ or at least an approximation to it. This can be achieved by the posterior updating approaches described above.

As an example, \citet{kue-etal:smc} use Shannon information as utility function for their SMC sampler. They refer to \citet{seb-wyn:sha} who show that under some assumptions such as stationarity
\begin{equation*}
\int_{\thetab \in \Theta} \int_{\zb \in \Y} \int_{\thetab' \in \Theta} p(\thetab) p(\zb|\thetab,\db) [p(\thetab'|\zb,\db) \log p(\thetab'|\zb,\db) ] d \thetab' d \zb d \thetab
\end{equation*}
is equal to the negative entropy of the marginal with respect to $\zb$,
\begin{equation*}
C - \int_{\zb \in \Y} p(\zb|\db) \log p(\zb|\db) d \zb,
\end{equation*}
which helps to reduce the complexity of the SMC sampler of \citet{kue-etal:smc} to some degree.

\section{Example}

\subsection{Example setting}

We apply the simulation-based design methodology developed in the previous sections to a standard linear regression example. In that case the likelihood function is of a well-known and simple form, so there is no need to invoke likelihood-free methods. The purpose of our example is merely to demonstrate various important aspects one has to consider when applying simulation-based design algorithms with likelihood-free extensions. For our example it is very easy to calculate the posterior distribution given previous observations, so it is possible to compare the likelihood-free approximations of the posterior distribution to the exact posterior distribution. Moreover, here the expected utility integral can also be computed analytically. This allows us to compare the results from the simulation-based optimal design algorithms to the exact results. Thus we are able to analyze the sensitivity of the result to the approximation quality and to what extent  approximation quality depends on the settings of the ABC algorithms.

We assume that
\begin{equation*}
\zb|\thetab,\db \sim \mathcal{N}(\Db \thetab, \sigma^2 \Ib_n).
\end{equation*}
That is, the expected value of the dependent variable is a linear combination of the parameter values $\thetab \in \Theta \subseteq \mathbb{R}^k$ and depends on the design through the design matrix $\Db = (\mathbf{f}(d_1),\ldots,\mathbf{f}(d_n))^T$, where $\mathbf{f}(.)$ is a $k$-dimensional function of the design variables $d_i \in [-1,1]$ and $\db = (d_1,\ldots,d_n)$. The $n$ observations are assumed to be normally distributed, independent, and homoscedastic with known variance $\sigma^2$.

We assume that $s$ previous observations $\yb = (y_1,\ldots,y_s)$ have been collected which follow the same distribution:
\begin{equation*}
\yb|\thetab,\bm{\xi} \sim \mathcal{N}(\Kb \thetab, \sigma^2 \Ib_s),
\end{equation*}
where $\Kb = (\mathbf{f}(\xi_1),\ldots,\mathbf{f}(\xi_s))^T$ and $\bm{\xi} = (\xi_1,\ldots,\xi_s)$.

Furthermore, the parameters $\thetab$ follow the prior normal distribution
\begin{equation*}
\thetab \sim \mathcal{N} (\thetab_0, \sigma^2 \Rb^{-1}).
\end{equation*}

The posterior distribution of $\thetab$ given the previous observations can easily be computed. Introducing $\Lb = \Kb^T \Kb$, the posterior is given by
\begin{equation*}
\thetab|\yb,\bm{\xi} \sim \mathcal{N} \left( (\Lb + \Rb)^{-1} (\Kb^T \yb + \Rb \thetab_0), \sigma^2 (\Lb + \Rb)^{-1} \right).
\end{equation*}
In the same way, the updated posterior distribution can be obtained as
\begin{equation*}
\thetab|\{\zb,\yb\},\{\db,\bm{\xi}\} \sim \mathcal{N} \left( (\Mb + \Lb + \Rb)^{-1} (\Db^T \zb + \Kb^T \yb + \Rb \thetab_0), \sigma^2 (\Mb + \Lb + \Rb)^{-1} \right),
\end{equation*}
where $\Mb = \Db^T \Db$.

We take $u(\zb,\db,\thetab) = \log \left( \frac{p(\thetab|\{\zb,\yb\},\{\db,\bm{\xi}\})}{p(\thetab)} \right)$ as our utility function, so that the expected utility for a specific design $\db$ is the expected gain in Shannon information (see \citealt{cha-ver:rev}):
\begin{eqnarray*}
U(\db) & = & \int_{\zb \in \mathbb{R}^n} \int_{\thetab \in \Theta} u(\zb,\db,\thetab) p(\zb,\thetab|\yb,\{\db,\bm{\xi}\}) d \thetab d \zb \\
% & = & \int \log \left( \frac{p(\theta|y,X)}{p(\theta)} \right) p(\theta|y,X) p(y|X) d \theta dy \\
 &=& \int_{\zb \in \mathbb{R}^n} \int_{\thetab \in \Theta} \log \left( \frac{p(\thetab|\{\zb,\yb\},\{\db,\bm{\xi}\})}{p(\thetab)} \right) p(\zb|\thetab,\db) p(\thetab|\yb,\bm{\xi}) d \thetab d \zb.
\end{eqnarray*}

Since $\int \int \log (p(\thetab))   p(\zb|\thetab,\db) p(\thetab|\yb,\bm{\xi}) d \thetab d \zb = \int \log (p(\thetab)) p(\thetab|\yb,\bm{\xi}) (\int p(\zb|\thetab,\db) d \zb) d \thetab $ does not depend on $\db$,
for optimization it would be sufficient to compute
\begin{eqnarray*}
U^{*}(\db) & = & \int_{\zb \in \mathbb{R}^n} \int_{\thetab \in \Theta} u^{*}(\zb,\thetab,\db)   p(\zb|\thetab,\db) p(\thetab|\yb,\bm{\xi}) d \thetab d \zb \\
 & = & \int_{\zb \in \mathbb{R}^n} \int_{\thetab \in \Theta} \log (p(\thetab|\{\zb,\yb\},\{\db,\bm{\xi}\}))  p(\zb|\thetab,\db) p(\thetab|\yb,\bm{\xi}) d \thetab d \zb.
\end{eqnarray*}

For our particular model, the integral can be computed analytically and is given as  $$U^{*}(\db) = -\frac{k}{2} \log(2 \pi) - \frac{k}{2} + \frac{1}{2} \log \det \left( \sigma^{-2} (\Mb + \Lb + \Rb) \right).$$
 It has the same maximum as the criterion for $D_B$ optimality, $\Psi(\db) = \det \left(\Mb + \Lb + \Rb \right)$ (cf. \citealt{atk-etal:opt}). Note that the $D_B$-optimal design does neither depend on $\sigma^2$ nor on the prior mean $\thetab_0$ nor on the previous observations $\yb$.

We choose a setting for which the exact solution can be obtained easily, and thus a comparison of  the results  of our design algorithms is feasible.

The following setting is used: the predictor is a polynomial of order two in one factor, i.e.
$$\Db = \left( \begin{array}{ccc}
1 & d_1 & d_1^2 \\
1 & d_2 & d_2^2 \\
\vdots & \vdots & \vdots \\
1 & d_n & d_n^2
\end{array} \right) \quad \text{and} \quad
\Kb = \left( \begin{array}{ccc}
1 & \xi_1 & \xi_1^2 \\
1 & \xi_2 & \xi_2^2 \\
\vdots & \vdots & \vdots \\
1 & \xi_s & \xi_s^2
\end{array} \right).$$

The continuous optimal design for this problem puts equal weights of 1/3 on the three design points $-1$, $0$, and $1$, see \citet{atk-etal:opt}. Likewise, if the number of trials of an exact design is divisible by three, then at the optimal design 1/3 of the trials are set to $-1$, $0$, and $1$, respectively. For our first example, we choose the prior information matrix $\Rb$ in a way so that it represents prior information equivalent to one trial taken at the design point $0$, i\@.e\@. $\Rb = \mathbf{f}(0) \mathbf{f}^T(0) = (1,0,0)^T (1,0,0)$. A value of $10^{-5}$ is added to the diagonal elements to make $\Rb$ invertible. Furthermore, we assume that one previous observation has been collected at the design point $-1$, so that $\Lb = \mathbf{f}(-1) \mathbf{f}^T(-1) = (1,-1,1)^T (1,-1,1)$. If we have $n=1$ (future) trial, it is optimal to set this trial to 1.

\subsection{MCMC sampler for augmented target distribution}

We applied two different design algorithms to this example. The first one is an MCMC sampler where the target distribution for the expected utility function is augmented with the set of simulated previous observations $\xb_{1:M,1:J}$. This is the first algorithm described in Section \ref{ABC_simulation_based_design}. We set the power of the expected utility function, $J$, to 1 and the number of repetitions of the simulated previous observations, $M$, also to 1. The design dimension of this problem is one, so there is no need to use a higher power $J$ or an increasing schedule for $J$. The mode of the sampled distribution can easily be read off from a histogram or a kernel density plot of the sample. As neighborhood kernel we use the uniform kernel: $$p_{\epsilon}(y,x) \propto \mathbb{I}_{|y-x|<\epsilon}(x).$$
We use the uniform distribution on the interval $[-1,1]$ as independence proposal distribution for $\db = d$. For our example this is a reasonable choice because the utility surface is rather flat. Furthermore, we set $\sigma^2 = 2$, $\thetab_0 = (0,0,0)^T$, and we assume that the previously collected observation at $\xi = -1$ is $y = 40$. Note  that these parameters should have no effect on the outcome in our example.

We have run the algorithm for various values of $\epsilon$ ($\sigma, 2 \sigma, 4 \sigma, 8 \sigma, 16 \sigma$) and for various lengths of the Markov chain ($10^4, 10^8, 10^9$). Due to memory allocation constraints the output of the Markov chains of length $10^8$ and $10^9$ was thinned, keeping every 10\textsuperscript{th} and 100\textsuperscript{th} element of the chain, respectively.

The utility function $u(z,d,\thetab)$ is not non-negative everywhere. If negative utilities occur, the simulation step is repeated until the sampled utility is positive. This modification distorts the output of the estimated utility surface, but we are only interested in regions of high expected utility anyway. If we use the utility function $u(z,d,\thetab) = \log(p(\thetab|\{z,y\},\{d,\xi\})) - \log(p(\thetab))$, we do not observe many cases with negative utilities. One could also add a positive constant to the utility function if too many negative utilities are sampled.

The differences between the outputs for the various Markov chain lengths are hardly surprising. Especially for low $\epsilon$, however, thinning does not reduce the integrated autocorrelation time (IAT) \footnote{The integrated autocorrelation time of a process is defined as $\text{IAT} = 1 + 2 \sum_{i=1}^{\infty} \rho_i$, where $\rho_i$ denotes the autocorrelation of the process at lag $i$. It roughly says that one observation drawn from a hypothetical i\@.i\@.d\@. sampler with the same stationary distribution as the MCMC sampler is worth IAT observations drawn from the MCMC sampler. We estimate the IAT using the estimating procedure proposed by \citet{gey:pra}.} proportionally. On the other hand, a Markov chain of length $10^4$ is often too short to properly represent the expected utility surface because of the very low acceptance rates. We therefore have found the Markov chain sampler with $10^8$ iterations to be more efficient than the sampler running for $10^9$ iterations while still giving very accurate results. On a PC with an Intel Core i3 CPU (2.10 GHz) and 4 GB RAM the MCMC samplers with $10^8$ iterations needed from 2.25 to 3 minutes to produce their sample.

Table \ref{Full_MCMC_vary_epsilon} shows the acceptance rates and IATs for MCMC samplers with $10^8$ iterations for different values of $\epsilon$. A higher accuracy for the estimation of the posterior distribution is clearly associated with a higher IAT. One has to find a reasonable compromise between these two conflicting goals.

\begin{table}[htbp]
\begin{center}
\caption{Acceptance rates and IATs for MCMC sampler on the augmented utility with $10^8$ iterations for different values of $\epsilon$. \label{Full_MCMC_vary_epsilon}}
\begin{tabular}{|D{.}{.}{2}|D{.}{.}{4}|D{.}{.}{2}|}
\hline
\multicolumn{1}{|c|}{$\epsilon$} & \multicolumn{1}{c|}{Accept. rate} & \multicolumn{1}{c|}{IAT} \\
\hline
1.41 & 0.0048 & 44.01 \\
2.82 & 0.0092 & 22.75 \\
5.64 & 0.0149 & 14.05 \\
11.28 & 0.0163 & 12.93 \\
22.56 & 0.0170 & 12.28 \\
\hline
\end{tabular}
\end{center}
\end{table}

\begin{figure}[hbt!]\centering
 \begin{tabular}{cc}
  \includegraphics[width=0.5\textwidth]{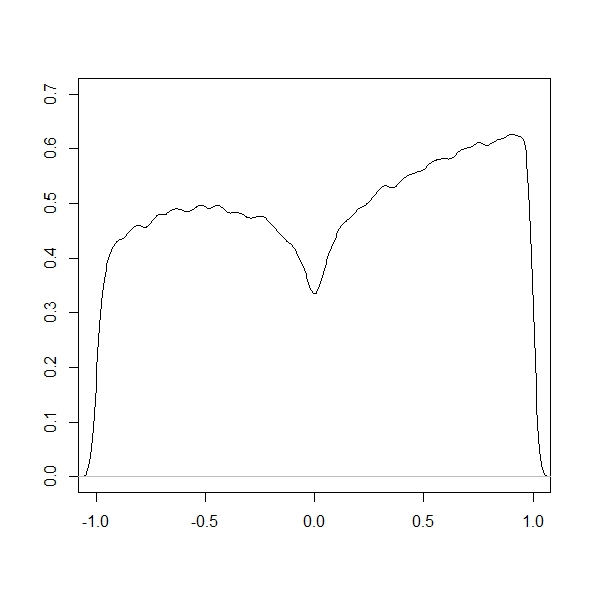} &
  \includegraphics[width=0.5\textwidth]{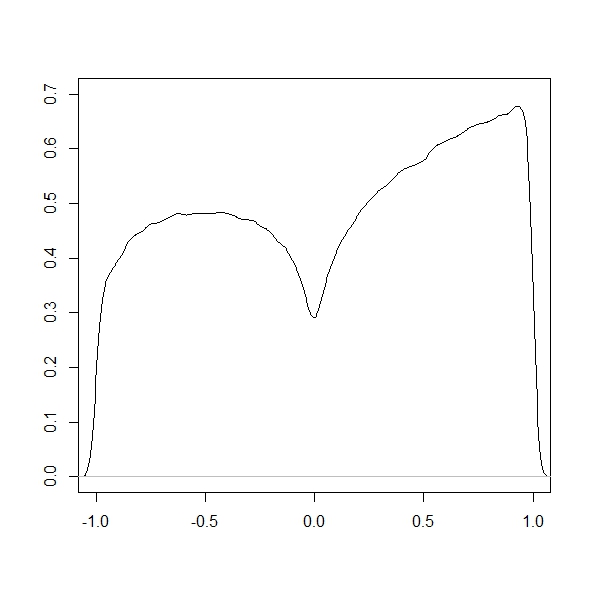} \\
\end{tabular}
\caption{Kernel density estimates of marginal distribution of $d$ for MCMC sampler on the augmented utility; $10^8$ iterations; \newline $\epsilon = 1.41$ (left) and $\epsilon = 22.56$ (right).\label{Full_MCMC_output}}
\end{figure}

 Figure \ref{Full_MCMC_output} suggests that the choice of $\epsilon$ has no effect on the result.  However, this is a special feature of our example and our choice of the utility function and not the case in general. In our example the value of $y$ does not matter for the optimal design, and hence it is irrelevant whether the simulated observations are close to the actual observations or not. Further, the utility is a function of the updated posterior. As the posterior is known explicitly,  the sampled values are just plugged in to compute the utility function. If  the posterior density function is not known,  one of the approaches outlined in Section \ref{utility_function} has to be pursued.  %There is no approximation uncertainty in evaluating the utility function.

We have also found that using a higher power $J$ proves problematic if this algorithm is used. In our example, the prior variance of $\thetab$ is very high due to the lack of prior information, so the prior distribution $p(\thetab)$ is very dissimilar to the posterior distribution $p(\thetab|\yb,\bm{\xi})$. Therefore, almost all simulated observations $\mathbf{x}_{m,j}$ fall outside some reasonable neighborhood of $\yb$ so that the rejection rates are very high. However, for the acceptance probability to be higher than zero, we have to have $\displaystyle \sum_{m=1}^M p_{\epsilon}(\yb,\xb_{m,j}) > 0 \quad \forall j \in \{1,\ldots,J\}$. Even if we take a high number $M$, this is very unlikely to happen for each $j = 1,\ldots,J$ if $J$ is sufficiently high. Thus, this algorithm is not suitable for higher-dimensional design problems where it is essential to have $J \gg 1$.

\subsection{Two-stage MCMC sampler}

The considerations mentioned above lead us to pursue a two-stage approach, where we first estimate the posterior distribution $p(\thetab|\yb,\bm{\xi})$ by the population Monte Carlo scheme suggested by \citet{bea-etal:ada}, and then use this approximate posterior in a standard MCMC simulation-based design algorithm as proposed by \citet{mue:sim}.

The population Monte Carlo scheme for approximate Bayesian computation works by creating a set of $N$ particles for which standard likelihood-free rejection sampling is applied. All particles have equal weights after the first step. Then in each consecutive step particles are created by sampling a particle from the previous step according to the weights, propagating this particle using a normal distribution with the mean set to the particle value at the previous step and the variance set to twice the empirical variance of the sample from the previous step. If the newly created particle is rejected, then a new particle is sampled from the particle set of the previous step and propagated. This is repeated until the particle gets accepted. This  procedure implements a form of rejection control: in the final sample there are only particles with non-zero weights.

In each step the target distribution changes. The idea is to start with a target distribution with a rather high $\epsilon$ so that the first sample is obtained quickly. In the subsequent steps $\epsilon$ is slowly getting smaller until the approximation of the posterior distribution is sufficiently accurate. The algorithm is of order $O(N^2)$ because for computing the weights one has to iterate over the whole particle set obtained at the previous step. Other algorithms like the one developed by \citet{del-etal:ada} only have complexity $O(N)$, but the population Monte Carlo sampler is rather reliable, unbiased, and easy to implement.

For the same setting as for the previous sampler we have computed the posterior distribution $p(\thetab|y=40,\xi=-1)$ using the population Monte Carlo method of \citet{bea-etal:ada}. The descending sequence of values for $\epsilon$ we used was $(16 \sigma, 8 \sigma, 4 \sigma, 2 \sigma, \sigma)$. The algorithm was run for particle sets of size 5000, 10000, and 20000. We know the exact posterior distribution, so we can compare the moments of the exact distribution to the moments of the weighted particle sample. The comparison of the mean, variance and covariance parameters shows that all different sample sizes deliver results that come very close to the exact posterior distribution, the deviations from the true values being less than 1 for the true mean parameters $(0.002,-19.998,19.998)$, for example. The deviations from the true variance and especially covariance parameters are more notable, but it makes no difference how large the particle set is. There is not much gain in having a particle set of size 20000, which took 13.71 minutes to compute, compared to a particle set of size 10000, which was computed in 3.41 minutes (the complexity is $O(N^2)$).

Since we have already obtained a weighted sample of the posterior distribution of size 20000, we use this sample as an approximation to the posterior distribution for the second stage. In the second stage we perform an ordinary MCMC simulation-based design algorithm as described by \citet{mue:sim}. We vary the power of the expected utility function ($J = 50, 100, 200$) and the length of the Markov chain ($10^5$ and $10^4$). We do not employ an increasing schedule for $J$ but have it set to a fixed value. The independence uniform proposal distribution for $d$ ensures that we still do not miss regions of high utility even if $J$ is quite high. We also performed the analyses for the case where we use the posterior sample of size 5000, but the results are not much different except for a slightly more uneven expected utility surface.

The length of the Markov chain has a rather small effect on the result, even though for high $J$ the acceptance rates are very low. In order to make sure that our sample is informative enough, however, we prefer a sample of size $10^5$, which took the computer 41 seconds to obtain.

\begin{figure}[hbt!]\centering
 \begin{tabular}{ccc}
  \includegraphics[width=0.3\textwidth]{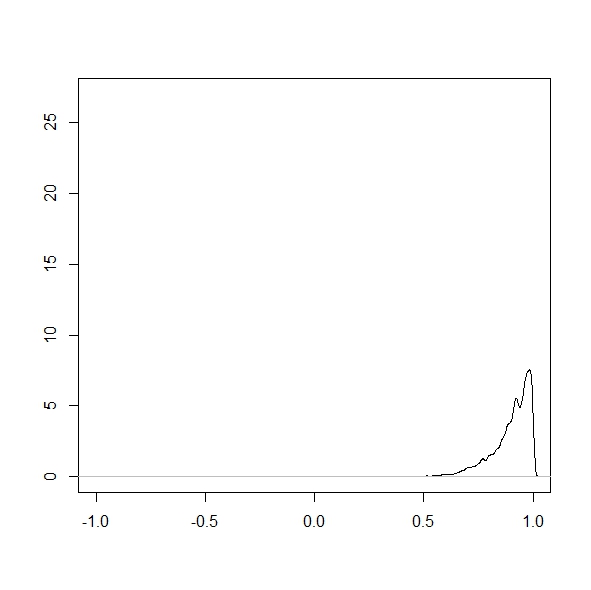} &
  \includegraphics[width=0.3\textwidth]{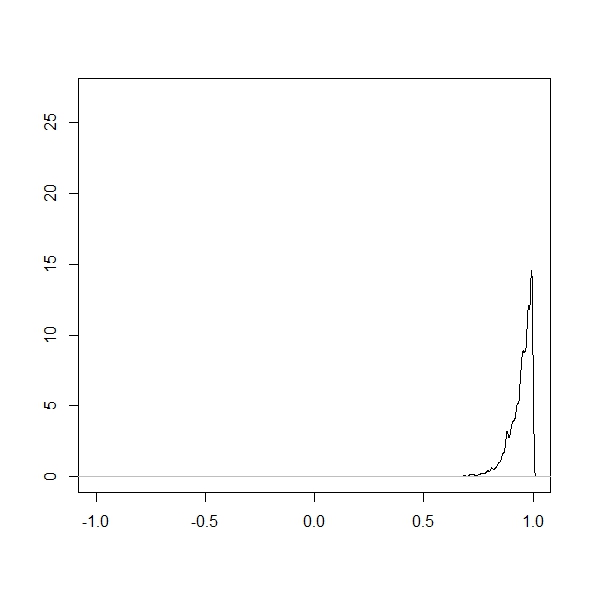} &
  \includegraphics[width=0.3\textwidth]{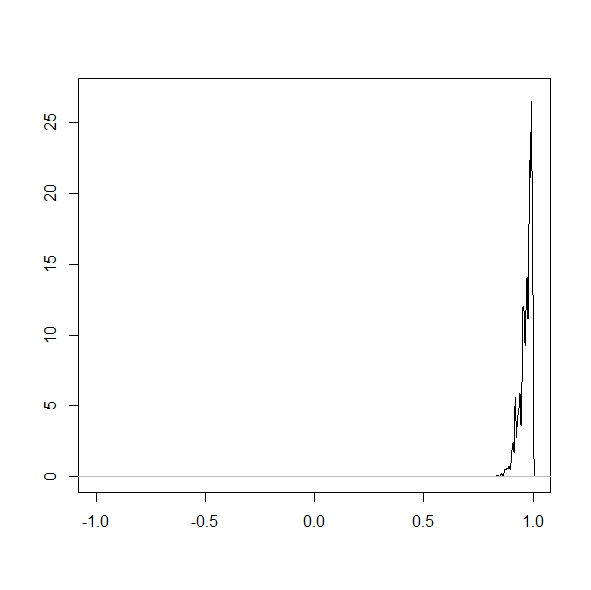} \\
\end{tabular}
\caption{Kernel density estimates of marginal distribution of $d$ for a two-stage MCMC sampler; one-dimensional design; sample length $10^5$; \newline $J = 50$ (left), $J = 100$ (center), and $J = 200$ (right).\label{Twostage_MCMC_output_dim1}}
\end{figure}

As would be expected, varying $J$ has the most pronounced effects on the outcome. Thus, Figure \ref{Twostage_MCMC_output_dim1} displays the kernel density estimates of the marginal sample of $d$ for $J = 50, 100,$ and $200$ when the Markov chain has length $10^5$. The corresponding acceptance rates and IATs are given in Table \ref{Twostage_MCMC_dim1}.

\begin{table}[htbp]
\begin{center}
\caption{Acceptance rates and IATs of a two-stage MCMC sampler for a one-dimensional design with sample length $10^5$ using different values of $J$. \label{Twostage_MCMC_dim1}}
\begin{tabular}{|r|D{.}{.}{4}|D{.}{.}{2}|}
\hline
\multicolumn{1}{|c|}{$J$} & \multicolumn{1}{c|}{Accept. rate} & \multicolumn{1}{c|}{IAT} \\
\hline
50 & 0.0628 & 38.96 \\
100 & 0.0260 & 108.13 \\
200 & 0.0078 & 524.14 \\
\hline
\end{tabular}
\end{center}
\end{table}

The two-stage approach can easily be extended to designs with more than one dimension. The setting is altered in the following way to create a two-dimensional design problem:\\$\Rb = \mathbf{f}(-1)\mathbf{f}(-1)^T + \mathbf{f}(0)\mathbf{f}(0)^T$ and $\bm{\xi} = (-1,1)^T$ so that the optimal design is either $\db = (0,1)^T$ or $\db = (1,0)^T$ (two maxima).
In the multidimensional case, the uniform ABC kernel function is defined as $$p_{\epsilon}(\yb,\xb) \propto \mathbb{I}_{||\yb-\xb||/\sqrt{n}<\epsilon}(\xb).$$
For the values of $\epsilon$ we choose the same sequence as in the one-dimensional case. We obtain posterior samples of size 20000 (13.39 min\@.), 10000 (3.35 min\@.), and 5000 (0.85 min\@.) for the posterior $p(\thetab | \yb = (-1,160)^T, \bm{\xi} = (-1,1)^T)$. The differences between the three samples are only marginal, all of them being very good approximations to the true posterior distribution. We use the sample of size 20000 for the second stage.

\begin{figure}[hbt!]\centering
 \begin{tabular}{cc}
  \includegraphics[width=0.4\textwidth]{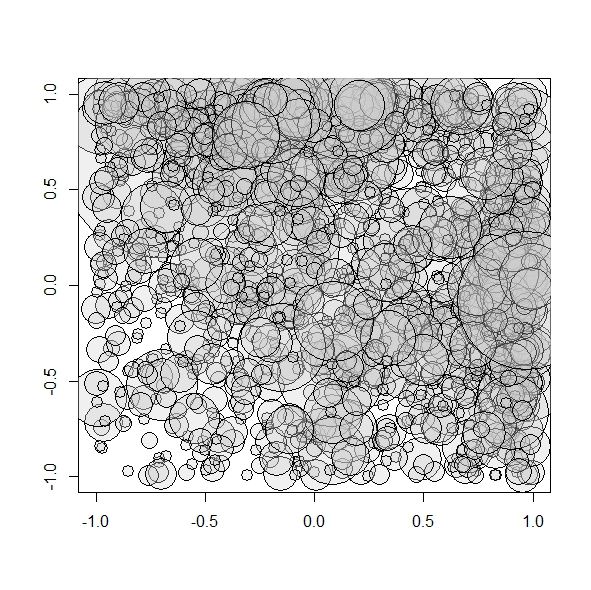} &
  \includegraphics[width=0.4\textwidth]{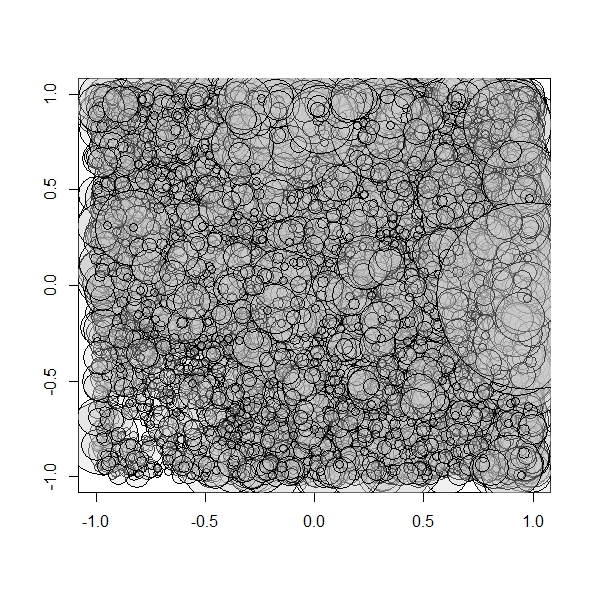} \\
  \includegraphics[width=0.4\textwidth]{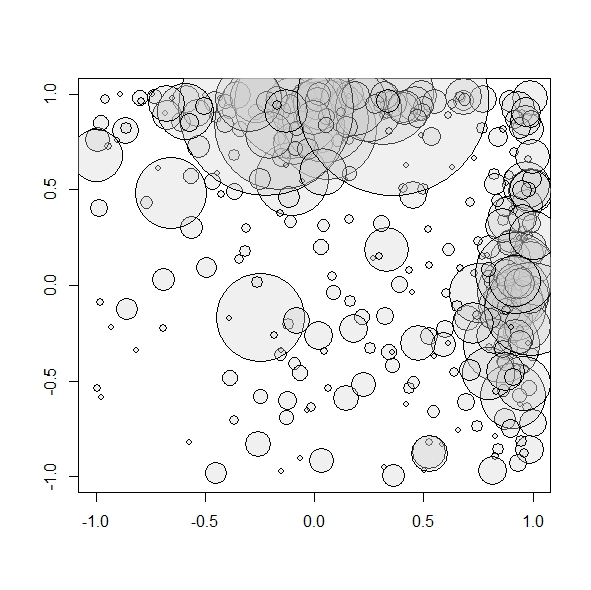} &
  \includegraphics[width=0.4\textwidth]{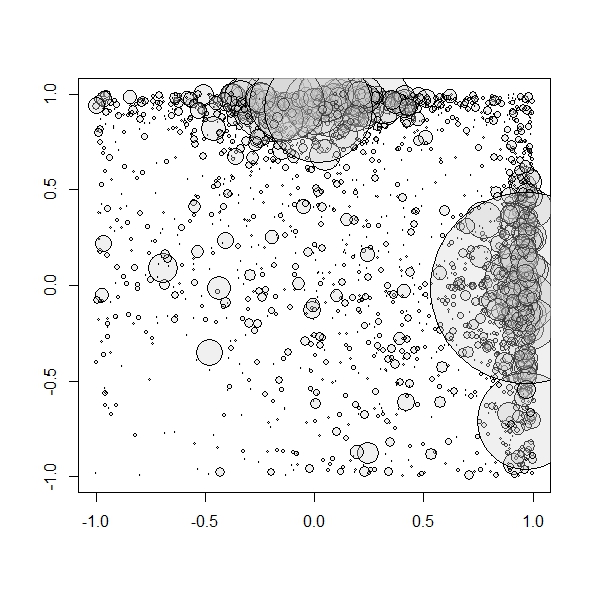} \\
  \includegraphics[width=0.4\textwidth]{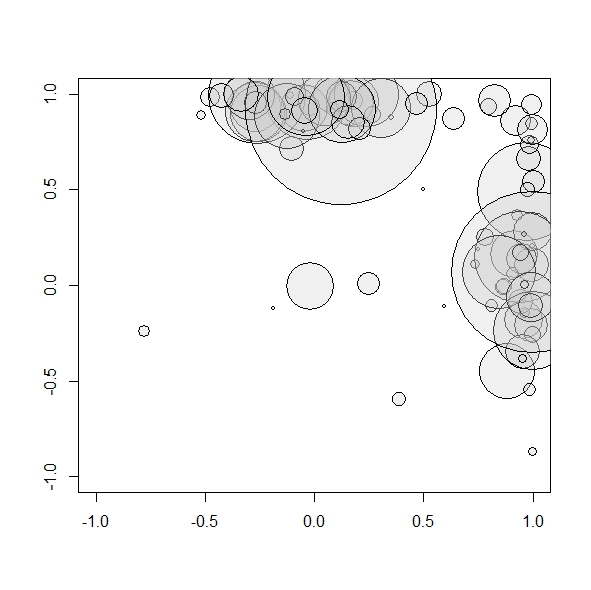} &
  \includegraphics[width=0.4\textwidth]{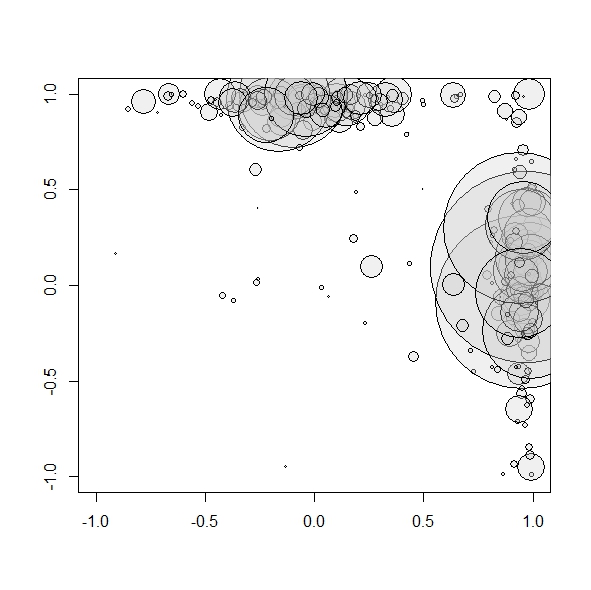} \\
\end{tabular}
\caption{Bubble plot of samples of marginal distribution of $\db$ (x-axis: $d_1$; y-axis: $d_2$) for a two-stage MCMC sampler; two-dimensional design; \newline sample lengths $10^4$ (left) and $10^5$ (right); $J = 50$ (top), $100$ (center), and $200$ (bottom); \newline circle areas proportional to the sample frequency of points (relative to the point with the highest frequency in the respective sample) . \label{Twostage_MCMC_output_dim2}}
\end{figure}

Figure \ref{Twostage_MCMC_output_dim2} shows the outputs of the samplers for $J = 50$, $J = 100$, and $J = 200$ and for the Markov chains of length $10^4$ and $10^5$. The longest runtime for the second stage is 1.19 minutes (sample length $10^5$ and $J = 200$), which is perfectly acceptable. One can see that for $J=200$ we get very good results. The sampled design points are very close to the optimal values for almost all iterations of the sampler.

\begin{table}[htbp]
\begin{center}
\caption{Acceptance rates and IATs of a two-stage MCMC sampler for a two-dimensional design using different values of $J$. \label{Twostage_MCMC_dim2}}
\begin{tabular}{|r!{\vrule width 1.5pt}D{.}{.}{4}|D{.}{.}{2}|D{.}{.}{2}!{\vrule width 1.5pt}D{.}{.}{4}|D{.}{.}{2}|D{.}{.}{2}|}
\hline
& \multicolumn{3}{c!{\vrule width 1.5pt}}{Markov chain length $10^4$} & \multicolumn{3}{c|}{Markov chain length $10^5$} \\
\cline{2-7}
\multicolumn{1}{|c!{\vrule width 1.5pt}}{$J$} & \multicolumn{1}{c|}{Accept. rate} & \multicolumn{1}{c|}{IAT $d_1$} & \multicolumn{1}{c!{\vrule width 1.5pt}}{IAT $d_2$} & \multicolumn{1}{c|}{Accept. rate} & \multicolumn{1}{c|}{IAT $d_1$} & \multicolumn{1}{c|}{IAT $d_2$} \\
\hline
50 & 0.1510 & 30.45 & 28.64 & 0.1411 & 42.29 & 43.79 \\
100 & 0.0352 & 88.55 & 149.71 & 0.0240 & 1109.95 & 1134.46 \\
200 & 0.0090 & 427.40 & 485.51 & 0.0026 & 2543.14 & 2457.598 \\
\hline
\end{tabular}
\end{center}
\end{table}

%\vskip 26.4  true pt
\vskip 15 true pt
\section*{Acknowledgements}
This work was partially supported by AANR-2011-IS01-001-01 DESIRE and FWF I 833-N18.

%\newpage

\bibliographystyle{abbrvnat}
\bibliography{sim_based_od_lit}

\end{document}